 \documentclass[logo,11pt,a4paper]{ETHpaper}

 \usepackage{amssymb,amsmath}           
 \usepackage{graphicx}                  
\usepackage{multirow}
\usepackage{wrapfig}
\usepackage[small,compact]{titlesec} 
\usepackage{natbib}
\usepackage{subfig}
\usepackage{color}
\usepackage[table]{xcolor}
\definecolor{BDPc}{HTML}{FFDD00}

\definecolor{Christianc}{HTML}{009BA8}

\definecolor{GLPc}{HTML}{ACC700}

\definecolor{Grunec}{HTML}{52A126}

\definecolor{FDPc}{HTML}{074EA1}

\definecolor{SPc}{HTML}{E52B38}

\definecolor{SVPc}{HTML}{7C7B80}

\definecolor{ALc}{HTML}{A31919}

\definecolor{Piratec}{HTML}{EB8A02}

\definecolor{NOALc}{HTML}{FFFFFF}

\definecolor{ao}{rgb}{0.0, 0.5, 0.0}
\definecolor{darkmagenta}{rgb}{0.55, 0.0, 0.55}
\definecolor{darkorange}{rgb}{0.8, 0.45, 0.0}

\author{David Garcia$^{1*}$ , Adiya Abisheva$^1$, Simon
  Schweighofer$^1$, Uwe Serd\"ult$^2$, Frank Schweitzer$^1$}

\authoralternative{D. Garcia$^{1*}$, A. Abisheva$^1$, S. Schweighofer$^1$, 
  U. Serd\"ult$^2$, F. Schweitzer$^1$}

\address{$^1$Chair of Systems Design, ETH  Zurich, Weinbergstrasse 56/58, 8092 Zurich, Switzerland\\
         $^2$Centre for Democracy Studies Aarau, Universit\"at Z\"urich,\\ K\"uttigerstrasse 21, 5000 Aarau, Switzerland\\
         $^*$Contact author: dgarcia@ethz.ch}

\date{\today}

\titlealternative{Ideological and Temporal Components of Network Polarization in Online Political Participatory Media. \href{http://onlinelibrary.wiley.com/doi/10.1002/poi3.82/abstract}{\textbf{Published in \textsl{Policy \& Internet}, \textbf{7(1)} (2015)}}}

\title{Ideological and Temporal Components of Network Polarization in Online Political Participatory Media}
\begin{document}
\mbox{}
\maketitle
\begin{abstract}

Political polarization is traditionally analyzed through the ideological
stances of groups and parties, but it also has a behavioral component that
manifests in the interactions between individuals.  We present an empirical
analysis of the digital traces of politicians in politnetz.ch, a Swiss online
platform focused on political activity, in which politicians interact by
creating support links, comments, and likes.   We analyze network
polarization as the level of intra-party cohesion with respect to inter-party
connectivity, finding that supports show a very strongly polarized structure
with respect to party alignment.  The analysis of this multiplex network shows
that each layer of interaction contains relevant information,  where comment
groups follow topics related to Swiss politics. Our analysis reveals that
polarization in the layer of likes evolves in time, increasing 
close to the federal elections of 2011. Furthermore, we analyze the internal
social network of each party through metrics related to hierarchical
structures, information efficiency, and social resilience.   Our results
suggest that the online social structure of a party is related to its
ideology, and reveal that the degree of connectivity across two parties
increases when they are close in the ideological space of a multi-party system.

\end{abstract}

\section{Introduction}

Political polarization is an important ingredient in the functioning of a
democratic system, but too much of it can lead to gridlock or even violent
conflict. An excess of political homogeneity, on the other hand, may render
democratic choice meaningless \citep{Sunstein2003}. It is therefore of
fundamental interest to understand the factors shifting this delicate balance
to one of the two extremes. Consequently, polarization has long been a central
topic for political science. Within this article, we build on this tradition,
but we also want to supplement it with several currently rather unregarded
aspects. First, large parts of political polarization literature are concerned
with two-party systems, particularly in the context of the United States
\citep{Waugh2009}. And second, polarization is generally conceptualized on
the basis of positions in ideological space \citep{Hetherington2009}. Commonly
used measures of political polarization are derived from those two premises,
and therefore conceptualize polarization as two ideological blocks (i.e.
parties) drifting apart on one political dimension, while increasing their
internal agreement.

Many democratic systems are characterized by more than two relevant parties,
making it problematic to apply standard ideology-based measures of
polarization to them \citep{Waugh2009}.   Besides methodological concerns, we
argue that an exclusive focus on ideological positions does not capture all
aspects of the term \emph{polarization}. It has been claimed that a sensible
definition of \emph{polarization} would have to comprise not only the
ideological stances of the polarized set of individuals or parties, but also
the interactions between them
\citep{Blau1977,Baldassarri2007,Conover2011,Gruzd2014}. At the same level of
ideological polarization, we would ascribe a higher level of  polarization to
a set of political actors when: i) they display a bias towards positive
interaction between actors of similar political positions, and ii) they tend
towards negative interaction between actors with dissimilar political
positions \citep{Guerra2013,Gruzd2014}. Assuming a feedback
between opinion and network polarization, a polarized society is divided
into by a small number of  groups with high internal consensus and sharp
disagreement between them \citep{Flache2011}. When establishing a social link
means to agree on opinions to certain extent \citep{Guerra2013}, \textbf{network
polarization} is defined as a phenomenon in which the underlying social
network of a society is composed of highly connected  subgroups with weak
inter-group connectivity \citep{Conover2012, Guerra2013}. The terms
\emph{clusters}, \emph{modules} or \emph{communities} are often used as
synonyms to define \textbf{groups} of individuals in a (social) network based
on their link topology. Such assignments are usually not unique, i.e.
individuals can be counted in different communities. Various algorithms for
network partitioning exist to optimize this assignment, often based on the
optimization of modularity metrics \citep{Newman2006}.
By adopting a network science
approach, we are able to capture and analyse the interaction aspect of
polarization, and at the same time to expand the study of polarization to
multi-party systems. In this work, we focus
on the network of support and interaction among \textit{politicians only}. 
This way, our analysis  is performed within
the context of \emph{elite polarization} \citep{Fiorina2008,Hetherington2009},
as a complementary view to previous studies of \emph{mass polarization} through
blogs and Twitter \citep{Guerra2013,Gruzd2014}.

Online political participatory media, like
\texttt{opencongress.org}\footnote{http://www.opencongress.org} and
\texttt{politnetz.ch}\footnote{http://www.politnetz.ch/}, serve as a digital
representation of a political system where voters and politicians can discuss
in an online medium. These political participatory media serve as
crowdsourcing platforms for the proposal and discussion of policies, leaving
digital traces that allow unprecedented quantitative analyses of political
interaction.  In this study, we present our analysis of \texttt{politnetz.ch},
a Swiss platform that allowed us to obtain data about online political participation.
On \texttt{politnetz.ch}, politicians and citizens freely discuss political topics, and also weave a social network
around them by expressing their support for politicians or by liking each
others' contributions. Though \texttt{politnetz.ch} is of course subject to
many limitations and distortions, it can be seen as a fairly faithful online
representation of Swiss politics. The plurality of Swiss politics allows us to
study polarization along party, as well as along ideological lines.
Additionally, the real-time quality of \texttt{politnetz.ch} data enables us
to analyze polarization with a very high temporal resolution. While
traditional research mainly focuses on changes in polarization within years or
decades \citep{Mccarty2006}, we can detect them within the scope of days and
weeks. This makes it possible to quantify the influence of day-to-day
political events, such as elections and referendums (which are of particular
relevance in Swiss politics \citep{Serdult2014}).

Previous works on the dynamics of polarization use agent-based modelling
approaches to simulate and analyze polarization in \emph{opinion} dynamics.
Here, we can distinguish between two different model classes. First, models
with \emph{binary} opinions, often called voter models, already imply that
these are opposite opinions. The question is then about the \emph{share} of
opposite opinions in a population of agents \citep{fs-voter-03}. Some
scenarios show the emergence of a majority favoring one opinion, or even the
convergence of the whole population toward the same opinion, called consensus.
A polarization scenario in binary opinion models results in the
\emph{coexistence} of opposite opinions, often with almost equal share.
Second, models with \emph{continuous} opinions focus on the (partial)
convergence of ``neighboring'' opinions such that groups of agents with the
same opinion emerge. Polarization in such models can occur if two of these
groups coexist, without any possibility to reach consensus
\citep{Groeber2009}. In most cases, opinion dynamics models assume an
interaction between  opinions and the underlying communication structure, with
agents of similar opinions communicating more frequently with each other than
with dissimilar agents. This can be a result of homophily (i.e., opinion
difference influencing structure) or of social influence (structure
influencing opinion). We will explore if those model assumptions also hold in
the empirical social network of \texttt{politnetz.ch}.

Online participatory media offer different possibilities of interaction
between individuals. Online actions, such as creating a social link or
commenting on a post, can be used in different context. Nevertheless,
interdependencies might exist between these interaction types. For example, if
users press the like button only to posts they positively comment, the
liking action would not contribute additional information to the
communication process.  In this work, one of our goals is to quantify  how
much information about one interaction type is contained in another, assessing
the added value of including all these interaction types in the analysis of
online communities.  In our analysis, we explore the three main interaction
types between politicians in \texttt{politnetz.ch}: \texttt{supports},
\texttt{likes}, and \texttt{comments}, to measure the differences of
politicians' behavior in each interaction context. For instance, do
politicians only \texttt{like} posts of politicians they \texttt{support}? Do
the patterns of \texttt{support} among politicians project onto
\texttt{comments} and \texttt{likes}?  In our analysis, we aim to  discover
groups of politicians determined by the networks of \texttt{supports},
\texttt{likes} and \texttt{comments}, and not necessarily by their party
affiliation.  It might be the case that two politicians that are the members
of the same party, might not support each other, and conversely, it can be the
case that two politicians that are \textit{not} the members of the same party
\textit{do} support each other. How often do such scenarios occur?  Does the
party affiliation of a politician define whom they \texttt{support},
\texttt{like}, and \texttt{comment}, producing network polarization?

Previous works centered around the US found strong polarization between left-
and right-leaning politicians \citep{Saunders2004}. However, left- and right-
leaning attitudes generally coincide with party membership in the US two-party
system. This way, it cannot be distinguished whether polarization is created
along the left-right dimension or along party lines, which can only be
differentiated in a multi-party system.  The US left- and right-aligned
communities also differ in their online interaction, as shown in previous
works using blogs \citep{Adamic2005}, \texttt{Twitter} \citep{Conover2012},
and \texttt{Youtube} \citep{Garcia2012}, opening the question of whether these
patterns prevail in other countries.  We analyze the network topology of
\texttt{politnetz.ch}, its evolution over time, and the topologies the social
networks within each party. As a consequence, we characterize the role of
elections and party ideology in network polarization, both at the party and at
the global level. The fact that most politicians in \texttt{politnetz.ch}
clearly state their party membership and make their online interaction with
other politicians via \texttt{likes}, \texttt{supports}, \texttt{comments}
publicly available, allows us to test the following hypotheses: i) political
polarization  is present in layers with a positive connotation, i.e.
\texttt{supports} and \texttt{likes}, ii) polarization, if present, is not
solely grounded upon political party alignment, but also depends on
politically relevant events, such as elections, and on the distance between
parties in ideological space.

Finally, we ask the question of whether the social structure of parties is
related to their ideology. Previous research in the US political system showed
differences in online network topology between the right- and left-aligned
groups \citep{Conover2012}. This opens questions whether i) there are
differences in the party structures of the political systems with more than two
parties, and whether ii) the pattern in the US for right- and left-
subcommunities holds in a multi-party system. We study Switzerland as an
example of a country with a multi-party political system, with the presence of
major and minor parties and additional ideological dimensions beyond
left and right.

\section{Materials and methods}
\label{ref:dataSec}

\subsection{Politnetz data}
\texttt{Politnetz.ch} is an online platform that enhances communication
among politicians and voters in Switzerland. Profiles in \texttt{politnetz.ch} 
are registered either for voters or politicians, where politician profiles
have an additional section called  ``Political information''.
This section includes the party a politician belongs
to, selected from among valid parties in
Switzerland\footnote{``Political information'' is exclusive to
  politician profiles and includes 4
  fields: party affiliation, political career, duration
  as a party member at the federal or local level, and
  involvement if any in Swiss "Verein", e.g. NPOs, NGOs, trade unions,
  or other organizations. The set of parties to choose from is limited to
  65 parties and politically relevant organizations in Switzerland. Politicians 
  are not able to insert other party names.}. 
Registration as a politician on \texttt{politnetz.ch} is not verified by the platform, 
due to the fact that politicians are not motivated to misrepresent
themselves. This becomes evident in their profile information: 
two thirds of the politician profiles include links to their homepages, 
while none of the voter profiles has an external website link. 

With respect to the amount of politicians in Switzerland,
\texttt{politnetz.ch} contains a large set of politician accounts, with a
certain bias towards German-speaking regions.  The amount of voter accounts is
more limited with respect to the size of the electorate, and voters do not
have any field for party affiliation in their profile. As mentioned above, 
we focus exclusively on politicians' accounts, excluding voters accounts from all analyses. 


Politicians have three major means of social interaction in
\texttt{politnetz.ch}: i) they can \texttt{support} other politicians, ii)
they can write posts and \texttt{comments} on the posts of other politicians,
and iii) they can explicitly \texttt{like} posts, which are publicly displayed
in their profile. Our \texttt{politnetz.ch} dataset includes the full track of
interaction between politicians for more than two years, including 3441
politicians, which created 16699 \texttt{support} links, 45627
\texttt{comments}, and 10839 \texttt{likes} to posts.  These digital traces
are analogous to the the data of previous works on \texttt{Twitter} for US
\citep{Conover2012} and German politicians \citep{Lietz2014}, where follower
relationships imitate \texttt{support} links, retweets resemble
\texttt{likes}, and mentions are used as \texttt{comments}.

Analyzing \texttt{politnetz.ch} data provides a complementary approach to
dominating \texttt{Twitter}-centric studies, which suffer of the \emph{model
organism} bias \citep{Tufekci2014}, in which  the prevalence of
\texttt{Twitter} as a data source can drive a scientific community to take
platform-dependent  findings as universal. Furthermore, \texttt{politnetz.ch}
is richer as a data source compared to \texttt{Twitter} for three reasons: i)
political alignment is explicit in and does not rely on external coding, ii)
digital traces include all three layers of interaction, in contrast with
\texttt{Twitter} data limitations that often  leave out the follower network
\citep{Conover2011, Aragon2013, Gruzd2014}, and iii) Politnetz provides
information about the creation of support links, in contrast with
\texttt{Twitter}'s opacity with respect to follower link creation times,
forcefully simplifying the follower network as static \citep{Conover2012,
Lietz2014}.

In our dataset, more than 80\% of the politicians declare themselves as
members of an existing party. We simplify the party affiliation data by
merging local and youth versions of the same party, creating  table in which
each politician is mapped to one of the 9 parties, or left unaligned. The
latter tag -- ``unaligned'' politician -- is an umbrella label for three
distinct categories of political affiliation: 1.3\% of the politicians stated
they are independent politicians\footnote{Germ. \textit{Parteilos},
http://www.parteifrei.ch/}, 2.9\% chose membership to non-governmental and
nonprofit organizations, and the remaining 13\% of the politicians who did not
provide affiliation information to any party or organization, which in general
are politicians active at the local level without alignment to any party at
the federal level.  In Table \ref{tab:statsParties} we show the absolute count
of politicians in the 9 parties and in the unaligned category.

\begin{table}[h!]
\centering
\begin{tabular}{|l|p{3cm}|c|c|c|}
\hline
Party& Description & N &Colour\\
\hline
SP & Social Democratic Party&609&\cellcolor{SPc}\\ \hline
SVP & Swiss People's Party, incl. EDU & 484&\cellcolor{SVPc}\\ \hline
FDP & The Liberals &473&\cellcolor{FDPc}\\ \hline
Christian & CSP, CVP, EVP &423&\cellcolor{Christianc}\\ \hline
Gr\"une & Green Party  &298&\cellcolor{Grunec}\\ \hline
GLP & Green Liberal Party &286&\cellcolor{GLPc}\\ \hline
\end{tabular}
\begin{tabular}{|l|p{3cm}|c|c|c|}
\hline
Party& Description & N &Colour\\
\hline
BDP & Conservative Democratic Party &151&\cellcolor{BDPc}\\ \hline
Piraten & Pirate Party &93&\cellcolor{Piratec}\\ \hline
AL & Alternative Left, incl. PdA &28&\cellcolor{ALc}\\ \hline
&Unaligned&596&\cellcolor{NOALc}\\\hline\hline
&Independent&47&\\
&NGO/Union&101&\\
&No affiliation&448&\\
\hline
\end{tabular}

\caption{Number of politicians in each of the 9 parties and in the unaligned category. Politicians
  labeled as unaligned can be further divided into three subgroups: 
  independent politicians, NGO/NPO/trade union
  representatives, and politicians with no party 
  affiliation. The fourth column shows colour codes of each of the 9 
  parties, which we use throughout this article.}
\label{tab:statsParties}
\end{table}

\subsection{Methods}

\begin{figure*}[ht]
\centering
\includegraphics[width=0.5\textwidth]{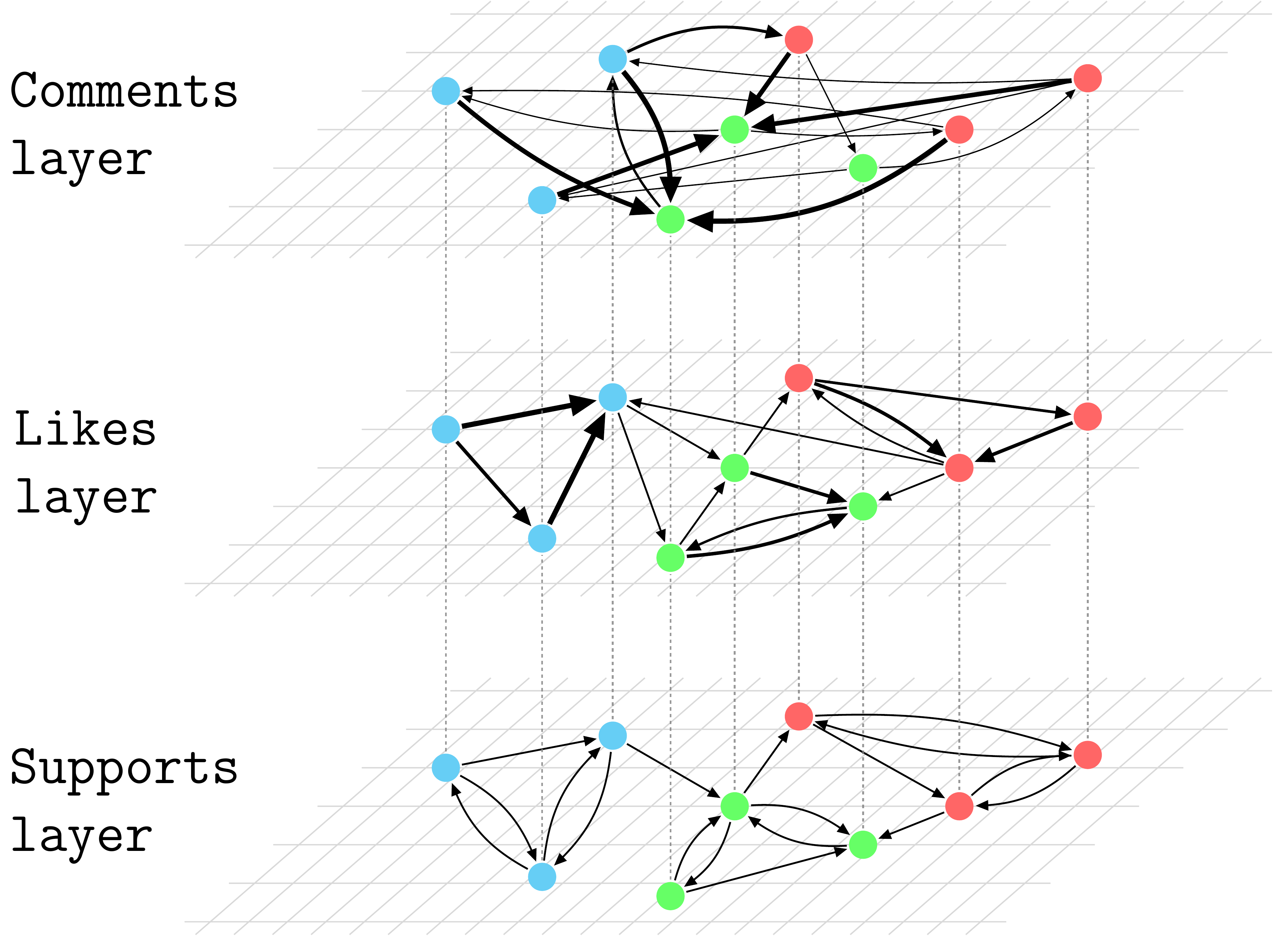}
\caption{Three layers of the \textit{multiplex} network in
  \texttt{Politnetz.ch}. The layer of \texttt{supports} is directed and
  unweighted, and the layers of \texttt{likes} and \texttt{comments}
are directed and weighted. Node colors illustrate an party affiliation, and link widths are proportional to their weights.}
\label{fig:multiplex}
\end{figure*}

\paragraph{Multiplex network analysis} The digital traces of politicians in
\texttt{politnetz.ch} allow us to study  interaction in three \textbf{network
layers}, one composed of \texttt{supports}, a second one of \texttt{likes},
and a third one of \texttt{comments}. Every node belongs to each layer, and
represents a politician with an account in \texttt{politnetz.ch}. A politician
$p_1$ has a directed link to another politician $p_2$ in the \texttt{supports}
layer if $p_1$ has $p_2$ in its list of supported politicians. The
\texttt{likes} and \texttt{comments} layers are also directed, but in
addition, links have weights equivalent to respectively the amount of
\texttt{likes} and \texttt{comments} that $p_1$ gave to the posts and comments
of $p_2$. These three layers compose a  \textbf{multiplex network}, also known
as a multimodal, multirelational or multivariate network,
\citep{Menichetti2013}, as depicted in Figure \ref{fig:multiplex}.  Multiplex
networks are a subset of multilayered networks:  Multiplex networks have one-
to-one relationships between nodes across layers and multilayered networks
have arbitrary connections across layers \citep{Boccaletti2014}. The
paradigmatic example of a multiplex network is a social network with different
types of social relationships (friendship, business, or family)
\citep{Szell2010}.   Examples of previously analyzed multiplex networks  are
air transportation networks, in which airports are connected through different
airlines, \citep{Cardillo2012}, and online videogames where players can fight,
trade, or communicate with each other, \citep{Szell2010}. 

\begin{figure*}[ht]
\centering
\subfloat[Supports layer, $|V| = 2,740$, $|E| = 16,657$][\centering Supports layer, \par $|V| = 2,740$, $|E| = 16,657$]{\includegraphics[width=0.33\textwidth]{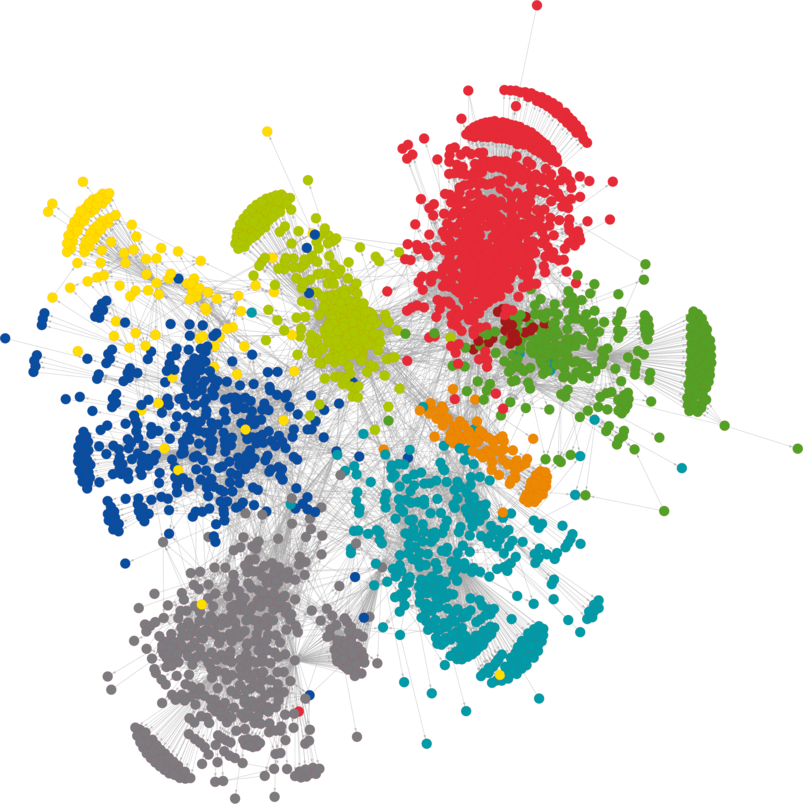} \label{fig:supports}}
\subfloat[Likes layer, $|V| = 1,147$, $|E| = 6,386$][\centering Likes layer, \par $|V| = 1,147$, $|E| = 6,386$]{\includegraphics[width=0.33\textwidth]{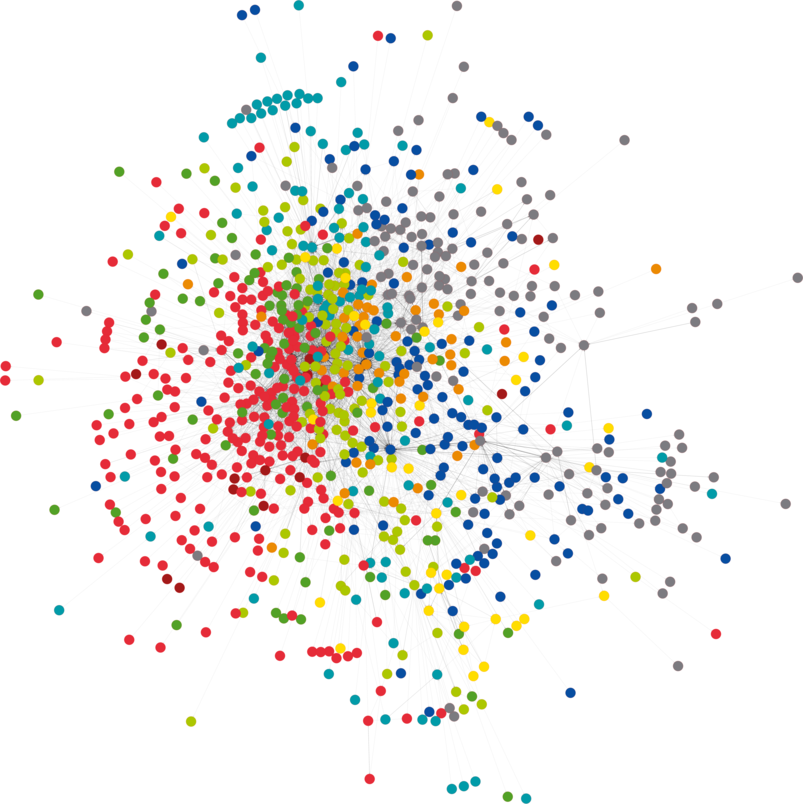} \label{fig:likes}}
\subfloat[Comments layer, $|V| = 1,091$, $|E| = 6,800$][\centering Comments layer, \par  $|V| = 1,091$, $|E| = 6,800$]{\includegraphics[width=0.33\textwidth]{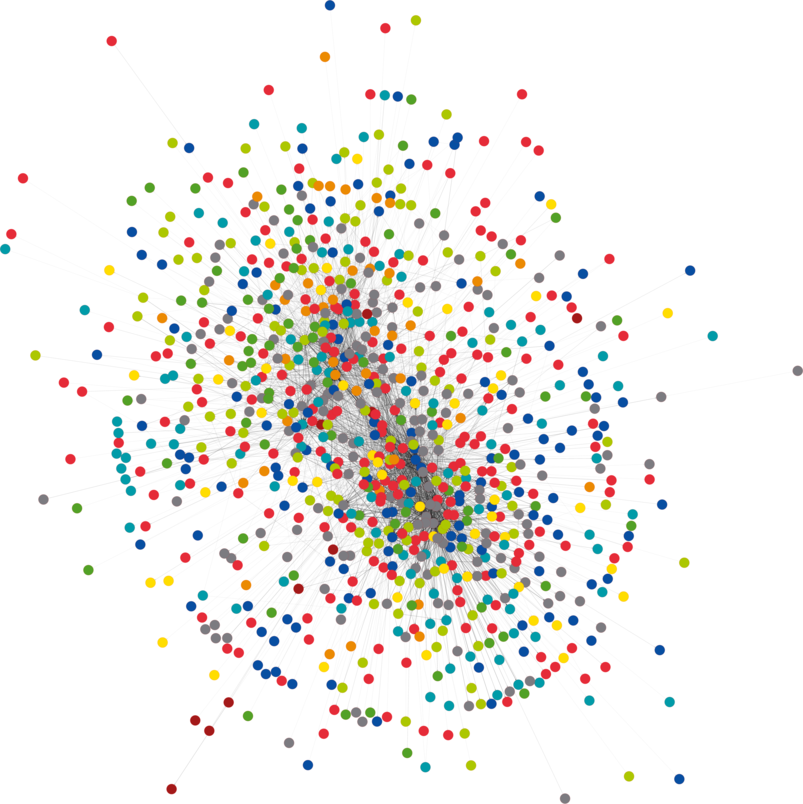} \label{fig:comments}}
\caption{Visualization of network layers of \texttt{supports}, \texttt{likes}
and \texttt{comments} excluding unaligned politicians. Colours of the nodes
are labelled according to the parties self-reported by politicians. Party
colors are reported in Table \ref{tab:statsParties}. The networks are drawn
using the Fruchterman-Reingold layout algorithm \citep{Fruchterman1991}.}
\label{fig:modViz}
\end{figure*}

The network layers are visualized in Figure \ref{fig:modViz}, where nodes
are colored according to their party alignment. Following the Fruchterman-
Reingold layout algorithm \citep{Fruchterman1991}, which locates connected
nodes closer to each other, an initial observation of Figure \ref{fig:modViz}
motivates our research question: politicians seem to be polarized along party
lines when creating \texttt{support} links, while this pattern is not so clear
for \texttt{likes} and \texttt{comments}.

In this multiplex network scenario, if a layer is ignored in the analysis,
relevant information about the interaction between nodes might be lost. By
measuring the interdependence of interaction types among politicians, it is
possible to address questions on the relationships across layers. These
relationships can exist at the link (microscopic) level, where links between
the same pairs of nodes co-occur in two different layers, or at the group
(mesoscopic) level, where the group to which each node belongs is similar
across layers.

To measure the overlap between layers at the link level,  we calculate the
\textbf{Partial Jaccard similarity coefficient} between the sets of links in two
layers. Given that the links of some layers are weighted, we calculate this
coefficient over binary versions of the weighted layers, i.e. in which all
weights are taken as 1. The Jaccard coefficient measures similarity in sets,
and is defined as the size of the intersection divided by the size of the
union of both sets.  It shows the tendency of nodes connected in one layer to also
connect in another layer (more details about its calculation can be found in
Appendix A.1).

The second metric we apply to both at the link and the group levels is the
\textbf{Normalized Mutual Information} (NMI) among layers.  The NMI of $X$
with respect to $Y$, NMI$(X|Y)$, is the mutual information (MI) between $X$
and $Y$  divided over the entropy of $X$ (see Appendix A.2). If the MI tells how much shared information
there is between $X$ and $Y$, and the entropy of $X$ quantifies information
content  of $X$ itself, then the NMI with respect to $Y$ measures to which
extent $Y$ contributes to information content of $X$. In other words, it gives
a fraction of information in $X$ that is attributed to $Y$, hence the NMI is
also known as the \textit{uncertainty coefficient} and has values between $0$
and $1$. With respect to $X$, if the NMI $\to 1$, then  knowing layer $Y$ can
predict most of the links in layer $X$; conversely, if the NMI $\to 0$, then
there is no dependence between the layers. To compute the NMI, we first
calculate information content of each matrix via the Shannon entropies with
the Miller-Madow correction. Then, we compute the empirical mutual
information between the layers, and finally obtain the normalized mutual
information with respect to each layer.  At the link level, the NMI of layer
$X$ with respect to layer $Y$ quantifies to which extent links in the network
layer $Y$ tell us about links in the network layer $X$. At the group level,
the NMI is computed over group labels, which allows the additional comparison
with the \emph{ground truth} of party affiliation, i.e. how much information
about party affiliation is contained in the  group structure  of certain layer
of interaction.

\paragraph{Network polarization}

Given a partition of politicians into groups, for example given their party
affiliation,  their network polarization can be computed through modularity
metrics.  We apply the  \textbf{Q-modularity} metric  \citep{Newman2006} to
each layer, measuring the tendency of politicians to link to other politicians
in the same group and to avoid politicians of other groups. The
\textbf{Q-modularity} of a certain partition of politicians in a layer has a
value between $-1$ and $1$, where $1$ implies that all links are within
groups, and $-1$ that all links are across groups.  This metric is specially
interesting, since it measures a comparison between the empirical network and
the ensemble of random networks with the same in-degree and out-degree
sequences,  allowing us to quantify the significance of our polarization
estimate against a null model (see Appendix B.1).

The natural partition of politicians is given by their party labels, from
which we obtain 9 groups and the \textit{unaligned} group as explained in
Table \ref{tab:statsParties}.  For each layer, we compute the network
polarization along party lines through the Q-modularity of party labels
$Q_{party}$. In addition, politicians can be partitioned in other groups
rather than parties, which could have higher modularity than  $Q_{party}$.
Finding such partition is a computationally expensive problem, which we
approximately solve by applying state-of-the-art community detection
algorithms, explained in Appendix B.2. This way, for
each layer we have another network polarization value $Q_{comp}$,
computationally found from the empirical data in that layer instead of from
party alignment.

To understand the origins of polarization, we analyze the time series
of network polarization over a sliding time window of two months,
taking into account only the links created within that window (see
Appendix B.3).  This allows us to empirically test if polarization
changes around politically relevant events, such as elections.
Furthermore, the comments layer includes contextual information in the
text of the comments between politicians of each group. To identify
the topics discussed in each group of the comments layer, we compute
the Pointwise Mutual Information (PMI) of each word in each group. The
PMI of a word compares its relative frequency within the group with
its frequency in all comments from all politicians, controlling for
significance as explained in Appendix C. This way, we produce lists of
words that highlight the discussion topics that characterize each
group of politicians in the comments layer.

\paragraph{Social structure and ideological position of parties}

Beyond network polarization, the multiplex network among politicians contains
information about the social structure of parties and the interaction of
politicians across parties. To complete our picture of Swiss online political
activity, we analyze the intra-party and inter-party structures present in the
network.

First, we apply metrics from social network analysis to compare the network
topologies inside each party, similarly to previous works on the US
\citep{Conover2012} and Spain \citep{Aragon2013}. With three types of social
interaction, we first have to select the layer that captures the social
structure of a political party in online participatory media. The
\texttt{supports} layer carries a positive connotation that does not change or
evolve in time. We choose to analyze \texttt{support} links, as leaders of the
party or politicians with authority will accumulate more \texttt{support}
links, but the amount of \texttt{likes} and \texttt{comments} they receive
depends on their activity in \texttt{politnetz.ch}. For each party, we extract
its internal network of supports, capturing a snapshot of its online social
structure in terms of leadership. We then calculate three network metrics to
estimate three structural properties of each network:

\begin{itemize} 

\item \textbf{Hierarchical structure: in-degree centralization.} The basic
idea of the network centralization  is to calculate the deviation of the in-
degree of each node from the \textit{most central} node, which has a special
position with respect to the rest in terms of influence \citep{Freeman1978}.
This way, in-degree centralization is computed as an average difference
between the in-degree of the politician with the most supports within the
party and the rest (see Appendix D.1). A
party with an in-degree centralization of 1 would look like a star in which
the central node attracts all supports, representing a network with the
strongest hierarchical structure. A party with in-degree centralization of 0
has support links distributed in a way such that every node has exactly the
same amount of supporters, showing the most egalitarian and least hierarchical
structure.

\item \textbf{Information efficiency: average path length.} This social
network metric measures the efficiency of information transport in a network:
shorter path length indicates an easily traversable network, in which it takes
fewer steps to reach any other node. The average path length is  defined as
the sum of the shortest paths between all pairs of nodes in a network
normalized over the number of all possible links in a network with the same
number of nodes, see Appendix D.2 for more details. We
compute the average path length between all pairs of connected nodes,  in
order to have a measure of the information efficiency of their social
structure.

\item \textbf{Social resilience: maximum \textit{k}-core.}  The ability of a
social group to withstand external stresses is known as \emph{social
resilience}. Social networks can display different levels of social resilience
from the point of view of cascades of nodes leaving the network, having a
resilient structure  if such cascades have small impact. Under the assumption
of rationality, the social resilience of a network can be measured through the
$k$-core decomposition \citep{Garcia2013}, indicating how many nodes will
remain under adverse conditions. This method assigns a $k$-core value to each
node by means of a pruning mechanism, explained more in detail in Appendix D.3. In essence, the $k$-core captures cohesive
regions of a network \citep{Seidman1983}, which are subsets characterized with
high connectedness, formally defined as the maximal subnetwork  in which all
nodes have a degree at least $k$.  Applying the $k$-core decomposition on the
subnetwork of each party, we aim at discovering such a resilient core of
political leaders, estimating the social resilience of a party as the maximum
$k$-core number of is social network. \end{itemize}


Second, we measure the level of inter-party connectivity by means of the
\textbf{demodularity} score, which measures the opposite of Q-modularity: the
tendency of a party to connect to another, as compared to a random ensemble of
networks. The demodularity score measures to which extent parties interact
with other parties, or how strongly politicians of one party preferentially
attach to politicians of \textit{another} party. This way, we compute a score
of demodularity from each party to each other party,  as explained more in
detail in Appendix E.

\paragraph{Party positions in ideological space} 

We quantify the ideological position of Swiss parties along the dimensions
\textit{Left-Right} and \textit{Conservative-Liberal} stance. This is
necessary to capture the multi-party system of Switzerland, in which the
position of parties cannot be simply mapped to a Left-Right dimension.  We use the party
scores of external surveys provided by \citet{Hermann2014}.  The authors of
the study give the following interpretation for both types of ideological
dimensions: the \textit{Left-Right} dimension expresses the understanding of
the concept of the state by each party. Left-wing politics sees the role of
the state in promotion of the economic and social well-being of citizens with
equally distributed welfare; while Right-wing politics understands the primary
role of the state as maintaining order and security\footnote{Additionally, the
difference   between the left- and right- politics comes with the difference
in   priorities set by each side. While the left-wing politics is   concerned
with the issues on the environment protection and asylum   granting policies;
the right-wing politics is committed to   strengthening of security forces, and to the competitiveness of the economy.}. The \textit{Conservative-
Liberal}  dimension encompasses the concepts of openness and willingness for
political changes. It covers the stance of parties on economics, social and political
issues, for instance, the position of a party on questions ranging from
globalisation to abortion\footnote{In the social area, political   issues such
as abortion, partnership law, etc. are covered; in the economic area,
questions of structural change, competition and attitude towards
globalisation, including reduction of subsidies,   free advertising, free
trade, etc., are discussed. And finally, in   the state policy field, debates
are between centralisation and   internationalization, such as Schengen and
international   peace-keeping missions, versus the preservation of the federal
system.}. These values are consistent with other sources of party positioning
data in Switzerland \citep{Germann2014}, stemming from Voting Advice Applications such as
\texttt{preferencematcher.org}\footnote{http://www.preferencematcher.org} and \texttt{smartvote.ch}\footnote{http://smartvote.ch}.

\section{Polarization in a multiplex network}
\label{sec:multiSec}

\subsection{Layer similarity}
\label{sec:linkNMISec}

Before measuring polarization in the three layers of the \texttt{politnetz.ch}
multiplex network, we need to verify if each layer contains additional
information, or if one can be predicted based on another.  Our first step is
to measure the similarity between layers at the link level, quantifying the
tendency of pairs of  nodes to connect in more than one layer. For this
analysis, we regard information in a network layer as expressed via presence
or absence of the interaction links between politicians, specifically we
extracted the binary, directed adjacency matrices of \texttt{supports},
\texttt{likes}, and \texttt{comments}.

We computed link overlaps between layers, as the ratio of links of layer
$\mathcal{X}$ that co-ocurr with links in $\mathcal{Y}$, among all links in
$\mathcal{Y}$.  We measure this overlap through the   partial Jaccard
coefficient of variables $X$ and $Y$,   which take value $1$ if certain link
is present in  layer $\mathcal{X}$ and in layer   $\mathcal{Y}$ respectively,
and $0$ otherwise (more details in Appendix A.1).
To test statistically the significance of these metrics, we apply the
jackknife bootstrapping method\footnote{\textit{Bootstrapping} is one type of the
resampling methods to assess accuracy (the bias and variance) of     estimates
or statistics (e.g. the mean or other metric in     study). \textit{Jackknife}
resampling is referred to obtaining a     new sample called \textit{bootstrap}
from the existing data by     leaving one observation out. With $N$
observations we obtain $N$     bootstrap samples. For each sample, we compute an
estimate,     in our case the Jaccard coefficient and the NMI. With
the set of $N$ statistics obtained from     the bootstrap samples, we are able to
calculate the standard deviation of our estimate.},  creating subsets of the network
obtained by leaving one node out. Overlaps across layers are reported in Table
\ref{tab:NMI}, revealing that the maximum overlap in the data is between
\texttt{comments} and \texttt{likes}, where 25.45\% of the \texttt{comments}
links have an associated link in the \texttt{likes} layer. While significantly
higher than zero, these values are relatively low, with more than 70\% of the
links not overlapping across layers.

\begin{table}[htb]
\centering
\begin{tabular}{|c | c | c | c | c | c | c|} 
 \hline
layer $\mathcal{X}$         & likes                 & supports                  & comments          & supports        & comments        & likes     \\ \hline
layer $\mathcal{Y}$         & supports                 & likes                  & supports          & comments        & likes         & comments    \\ \hline
overlap   & 18.13\%       & 6.96\%         & 7.13\%       & 2.91\%            & 23.9\%      & 25.45\%   \\
NMI       & 8.5\%             & 3.7\%              & 2.5\%        & 1.2\%       & 15.2\%      & 16\%    \\ \hline 
\end{tabular} 
 \caption{Link overlap and NMI across layers. Each measure was computed over
jackknife bootstrap estimates on each node, giving values of $2\sigma<
10^{-3}$.}
\label{tab:NMI} 
\end{table}

In addition, we computed the Normalized Mutual Information (NMI) between every
pair of the layers which estimates how much information of one layer is
contained in the links of the other one. The NMI of layer $\mathcal{X}$ with respect to
$\mathcal{X}$ tells us which fraction of information of the layer $\mathcal{X}$ is attributed to
knowing the layer $\mathcal{Y}$, as explained more in detail in Appendix
A.2. Table \ref{tab:NMI} shows the statistics for
the NMI between the three layers. All values are significantly larger than
$0$, and reveal weak correlations between the layers. 8.5\% of the information
in the \texttt{likes} layer is contained in the \texttt{supports}
layer, and less than 4\% of the information in the \texttt{supports} layer
is contained in the \texttt{likes} layer. \texttt{Supports} give 2.5\% of
the information in the \texttt{comments} binary network, and only 1.2\% is
contained in the opposite direction. The \texttt{likes} and \texttt{comments}
layers share normalized information of about 16\%, showing that there is
a bit of information shared across layers, but that they greatly differ in
most of their variance, in particular for the supports layer. These low levels
of overlap and NMI indicate that  each layer contains independent information
content that does not trivially simplify within a collapsed version of the
network.

\subsection{Network polarization}
\label{sec:polarSec}

The visualization of the three layers of the network in Figure
\ref{fig:modViz} suggests the existence of network polarization in the
\texttt{supports} layer, where politicians of the same party appear close to
each other. We quantify the level of network polarization among politicians in
each layer, given two types of partitions: i) by their party  affiliation, producing the $Q_{\text{party}}$
modularity score, and ii) by the groups found through computational methods on
the empirical data of each layer, resulting in the modularity score
$Q_{\text{comp}}$. Naturally, low modularity scores ($-0.5 \leq Q <
0.3$) imply that no polarization exists among politicians; high modularity
scores  ($0.3 \leq Q < 1$) will indicate the existence of polarization. Similar
values of  $Q_{\text{party}}$ and $Q_{\text{comp}}$ indicate that  the maximal
partition in a layer is close to the ground truth of party affiliation,  while
different values suggest that groups in a layer are not created due to
polarization along party lines. Across network layers, we
hypothesize that $Q_{\text{party}}$ is strong in the layers with the
positive semantics such as the links of \texttt{supports} or \texttt{likes}.
Furthermore, we investigate the role of the unaligned politicians by measuring
the network modularity including and excluding unaligned politicians. We aim
at testing whether these nodes act as cross- border nodes between parties,
therefore decreasing polarization when included in the analysis. For each
measure of polarization, we test its statistical significance by applying the
jackknife bootstrapping test on the networks -- by recomputing the modularity
score on the bootstraps -- each time with one node left out.

\begin{table}[ht]
\begin{tabular}{|l|c|c|c|c|c|c|c|c|c|}
\hline
&\multicolumn{3}{c|}{Supports}&\multicolumn{3}{c|}{Likes}&\multicolumn{3}{c|}{Comments}\\
\cline{2-10}
&$\left \langle Q \right \rangle$&$2\sigma$&|C|&$\left
  \langle Q\right \rangle $&$2\sigma$&$|C|$&$\left \langle Q\right \rangle$&$2\sigma$&$|C|$\\
\hline
$Q_{\text{party}}$ (incl.)&0.677&$6.8\cdot10^{-4}$&10&0.303&$11\cdot10^{-4}$&10&-0.009&$7.6\cdot10^{-4}$&10\\
$Q_{\text{comp}}$ (incl.)&0.746&$6.1\cdot10^{-4}$&13&0.460&$21\cdot10^{-4}$&12&0.341&$26\cdot10^{-4}$&16\\
$Q_{\text{party}}$ (excl.)&0.743&$6.1\cdot10^{-4}$&9&0.377&$13\cdot10^{-4}$&9&-0.009&$13\cdot10^{-4}$&9\\
$Q_{\text{comp}}$ (excl.) &0.745&$6.1\cdot10^{-4}$&10&0.472&$25\cdot10^{-4}$&17&0.336&$41\cdot10^{-4}$&16\\
\hline
\end{tabular}

\caption{Modularity score and number of groups found by the    community
detection algorithms and given politicians' parties labels   for the layers
including (first two rows) and excluding the unaligned   politicians. Standard
deviations are calculated through the jackknife   bootstrapping on nodes.  We
report the mean of modularity for bootstrap estimates, $\langle Q \rangle$.}

\label{tab:mod}
\end{table}



From the results in Table \ref{tab:mod}, we observe that the modularity score
slightly differs when unaligned politicians are ignored, having lower
polarization when they are present. This indicates that the unaligned group is
not cohesive, as it does not represent an explicit party, and unaligned
politicians connect across parties. For this reason, we remove unaligned
politicians from our subsequent analysis of polarization, as their absence of
affiliation does not signal their belonging to an additional group.

Among the three types of interactions, the \texttt{supports} layer shows the
highest modularity score $0.74$ for both $Q_{\text{party}}$ and
$Q_{\text{comp}}$, showing that, when making a \texttt{support} link,
politicians act as partisans. The \texttt{likes} layer is also polarized, but
the modularity score $0.47$ is lower than in the \texttt{supports} layer.
Hence, liking a post is still a signal of an adherence to a party, however
cross-party \texttt{like} links are more frequent  in comparison to cross-party
\texttt{support} links. Finally, the \texttt{comments} layer divided
algorithmically hints on a modular structure of the network resulting in
$Q_{\text{comp}}= 0.34$, however, such partition is not attributed to party
membership of politicians, with $Q_{\text{party}}=-0.009$. This suggests that
\texttt{comments} group politicians around discussion topics, motivating our
investigation of the origins of polarization in the layer of \texttt{comments}
in Section \ref{sec:originsSec}.

\subsection{Group similarity across layers}
\label{sec:commNMISec}

The similar values of $Q_{party}$ and $Q_{comp}$ for the supports layer
suggest that the partition of politicians into parties might be very similar
to the results of community detection algorithms, while the different values
for comments suggest the opposite. To empirically test this hypothesis,   we
compare group and party labels among politicians in each layer and across
layers through the NMI at the group level.  Within each layer, the NMI tells
whether  group labels discovered algorithmically can be predicted via party
labels and vice versa. To do this, we follow a similar methodology as in
Subsection \ref{sec:linkNMISec}. We compute the entropies of a layer (see Appendix A.2) based on
detected groups and based on party labels, then we calculate the mutual
information between the different partitions, and finally obtain the NMI scores
with respect to algorithmic and party partition. In this application of the
NMI, random variables are the  group labels of each node in each layer and
party affiliation.

The results in Table \ref{tab:NMIcommunity} show  that \texttt{supports}
groups  and party labels mostly match, $84-90\%$, contrary to \texttt{likes}
and \texttt{comments}. Across layers, only \texttt{comments} and
\texttt{likes} show a weak similarity in group labels $8-11\%$ and nearly no
similarity to the \texttt{supports} layer. This result allows us to confirm
that polarization in \texttt{supports} is due to party alignment of
politicians, which does not hold for \texttt{likes} and \texttt{comments}.
The high modularity score in the \texttt{likes} layer and the low information overlap
with the \texttt{comments} layer tells us that the \texttt{like} signal has a
dichotomous role as a social link. In certain situations, a \texttt{like} to a
post signals a party affiliation; in other scenarios, it also shows
politicians favour posts not only based on a party membership but due to the
content of the post.

\begin{table}[ht]
\centering
\begin{tabular}{|c | c | c | c | c | c | c|} 
 \hline
X         & parties   &  parties    &  parties     &  supports     &  supports   &  comments  \\
Y         & supports  &  likes      &  comments    &  likes        &  comments   &  likes     \\
\hline
NMI(Y|X)  &  90.04\%  &  3.99\%     &  3.29\%	   &  3.83\%	   &  3.4\%		 &  11.77\%   \\
NMI(X|Y)  &  84.51\%  &  3.87\%     &  4.4\%	   &  4.12\%	   &  4.5\%		 &  8.56\%    \\
 \hline
\end{tabular}
\caption{Normalized mutual information of the group labels
  computationally found for each layer, and the party
  label. Groups of \texttt{supports} 
are very similar to parties, but the rest has low mutual information.} 
\label{tab:NMIcommunity}
\end{table}


\section{Origins of network polarization} 
\label{sec:originsSec}

\subsection{Topic analysis of comment groups}
\label{sec:wordcloudSec}

The above results show the existence of a partition of politicians in the
\texttt{comment} layer that conveys certain polarization
($Q_{\text{comp}}=0.336$), but which does not match to their party alignment.
This suggests that \texttt{comments} happen more often across parties,
partitioning politicians into groups by some other property besides party
affiliation. To understand the reasons for such partition, we
investigate the content of the \texttt{comments} between the politicians of
each group. 10 of the 15 groups are very small -- they contain
reduced sets of politicians that exchange very few \texttt{comments} and are
isolated from the rest.  The 5 largest groups, which cover most of the
politicians,  have sufficient \texttt{comments} to allow us an analysis of
their words.  For each group, we computed a vector of word frequencies,
ignoring German stopwords\footnote{http://solariz.de/649/deutsche-
stopwords.htm}. To measure the extent to which a word is characteristic for a
group of politicians, we compute the \emph{Pointwise Mutual Information} (PMI)
of the frequency of the word in the group, compared to the frequency of the
word in the set of all \texttt{comments} (more details in Appendix C). We quantify the significance of the PMI through a
ratio test, only selecting words with $p<0.01$, producing the word lists
reported in Table \ref{tab:WordPMI}.

\begin{table}[ht]
\centering
\begin{tabular}{|c | c | c | c | c | c |} 
 \hline  
 			& C1  										& C2 								& C3 											&  C4 										& C5 							\\
 \hline 
 		1   & \color{darkmagenta}{ZauggGraf}			& \color{red}{SA} 					& \color{darkmagenta}{Noser} 					& \color{gray}{Fatah}						& Murmeln						\\
 		2   & Standardsprache							& \color{darkorange}{Hilton} 		& M\"angel 										& \color{gray}{Abbas}						& wolf		\\
  		3   & \color{red}{GfSt}							& \color{darkorange}{Dragovic} 		& Noten 										& \color{gray}{PA} 							& unsre							\\
  		4   & \color{darkmagenta}{Fitze}				& \color{blue}{Botschaftsasyl} 		& \color{red}{PK}	 							& \color{gray}{pal}							& \color{darkmagenta}{Lei}		\\
   		5   & \color{ao}{Digitalpolitik}				& \color{darkorange}{Hollande} 		& Raucher 										& \color{gray}{Hamas}						& 1291							\\
  		6   & \color{red}{\small Berufsbildungsfonds}	& \color{darkmagenta}{Spahr} 		& \color{red}{Weiterbildung} 					& \color{gray}{Libanon}						& Dokument						\\
 		7   & \color{ao}{Bahnpolizei}					& Jungsozialisten 					& \color{blue}{Asylsuchenden} 					& \color{gray}{Gaza}						& \color{darkmagenta}{Wolf}		\\
  		8   & \color{ao}{Kamera}						& \color{ao}{P21} 					& \color{darkmagenta}{{\small WidmerSchlumpf }}	& \color{gray}{Jordanien}					& Atomm\"ull					\\
 		9   & \color{darkmagenta}{Hannes}				& \color{darkmagenta}{Affentranger} & \color{red}{Pensionskassen} 	 				& \color{gray}{Westbank}					& einwenig						\\
  		10  & \color{darkmagenta}{Jeanneret}			& fur 								& \color{red}{Liberalisierung}					& \color{gray}{\small pal\"astinensischer}	& \color{ao}{Gripen}			\\
\hline 
 	words   & 356										& 207 								& 28 	 										& 129  										& 30							\\
   	comm.   & 10145										& 11312 							& 452  											& 899										& 656 							\\
    users 	& 436										& 273  								& 64	 										& 61 										& 51							\\
\hline
\end{tabular}
\caption{Top words for each of the 5 largest \texttt{comments} groups,
  ordered by PMI. Number of words with PMI significant at the 99\%
  confidence level, and amounts of users and comments in the
  group. Word colors classify them as follows: purple for Swiss
  Politician names, red for economic terms, blue for terms related to
  immigration, gray for words related to
  the conflict in the Middle East, green for terms related to
  security issues, and black for the rest. The top words significantly differ across
  groups, indicating that the partition in \texttt{comments} is aligned on
  the topics of discussion.} 
\label{tab:WordPMI}
\end{table}

All five groups have words with significant PMI, showing that they can be
differentiated from other groups with respect to the words that politicians
used in their \texttt{comments}.  This difference highlights the topics
discussed in every group, showing that the group structure in the
\texttt{comments} layer is driven by topical interests. While some
straightforward topics can be observed, especially for C4, we refrain from
interpreting the terms of Table \ref{tab:WordPMI} within the Swiss politics
context.  These results show that modularity in the \texttt{comments} layer
is topic-driven, and not party-driven, grouping politicians along their
interests and competences.

\subsection{The temporal component of polarization}
\label{sec:timeSec}

Our approach to polarization allows its measurement through behavioral traces,
which lets us create real-time estimates with high resolution.  The
\texttt{politnetz.ch} dataset provides timing information for the creation of
each \texttt{support}, \texttt{like}, and \texttt{comment}, giving us the
opportunity to study the evolution of polarization through time. Including a
time component in our analysis has the potential to reveal periods with higher
and lower polarization, allowing us to detect potential sources that create
polarization. We construct a set of  time series of polarization along party lines
$Q_{party}(t)$ in all three layers with a sliding window of two months, as
explained in Appendix B.3.

\begin{figure*}[h!]
\vspace{-20pt}
\centering
\includegraphics[width=\textwidth]{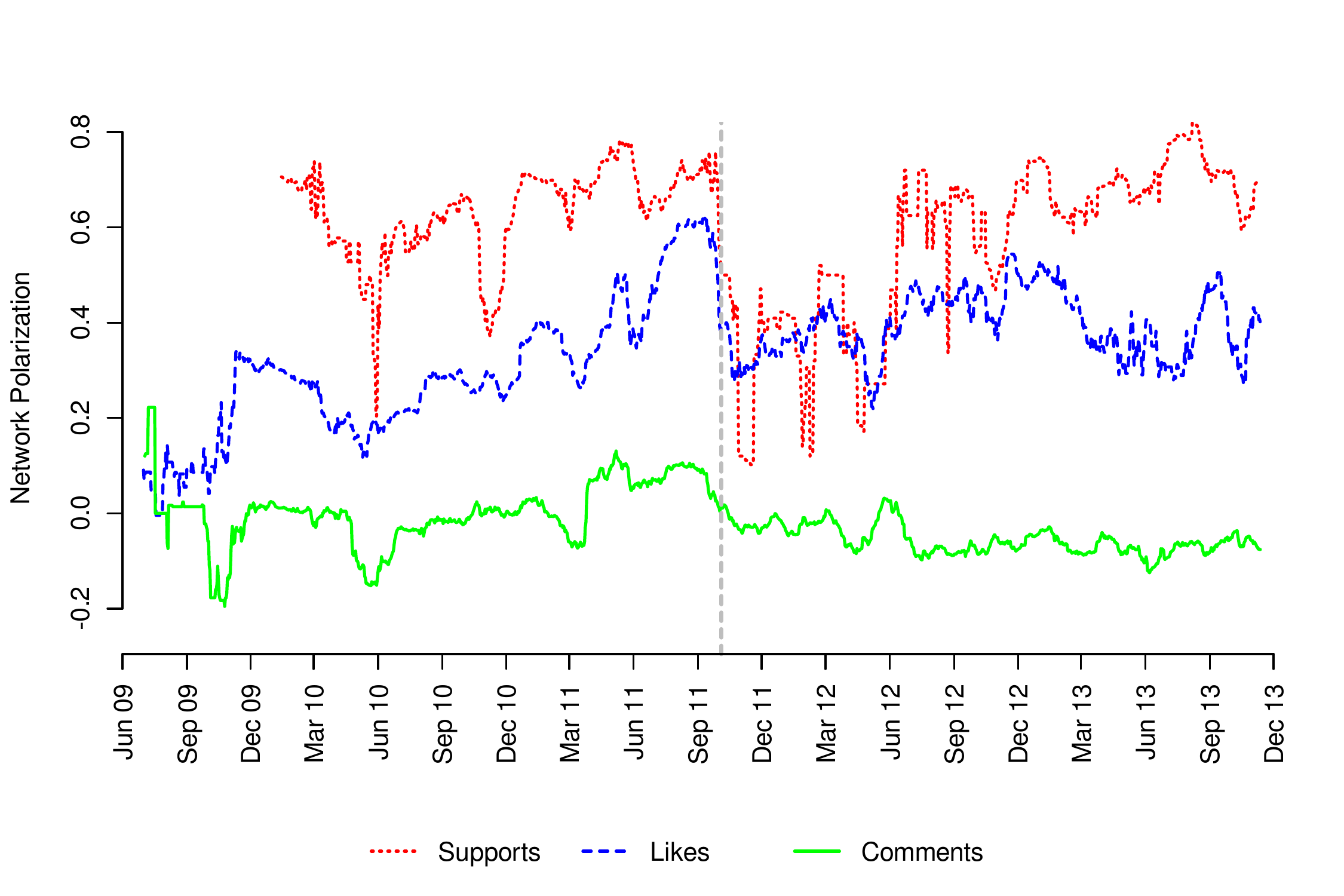} \label{fig:mod_overlay}
\vspace{-20pt}

\caption{Time series of Q-modularity by party labels of the
layers of \texttt{supports} (red), \texttt{likes} (blue) and
\texttt{comments} (green). The grey dashed line denotes Swiss federal
parliament election in 2011.  Two trends can be
observed: polarization among politicians peaks at pre-election time in the
network of \texttt{likes}, and post-election time is characterized by lower
levels of polarization in \texttt{likes} and \texttt{comments}.}

\label{fig:modTime}
\end{figure*}

Figure \ref{fig:modTime} shows the time series of network polarization along
party lines for the three layers of the network.  The   \texttt{comments}
layer shows negligible levels of polarization, fluctuating around $0$ for the
whole time period. Polarization in \texttt{likes} shows an increasing pattern
up to late 2011, reaching a value above 0.6 shortly before  the  Swiss federal
parliament election in 2011. Right after the election, polarization in
\texttt{likes} strongly corrects to levels slightly above $0.2$. Similarly,
polarization in \texttt{supports} has relatively stable values around $0.6$
before the federal election, dropping to values below $0.3$ right afterwards.
This analysis shows that network polarization, as portrayed in the online
activity of politicians, is not a stable property of a political system,
revealing that politically relevant events bias politician  behavior in two
ways: polarization reaches maxima during campaign periods, and polarization
levels relax quickly after elections.

The pattern in \texttt{likes} suggests that politicians avoid awarding a
\texttt{like} to a politician of another party when elections are close, but
in other periods they display a less polarized pattern that allows
\texttt{likes} across parties. On the other hand, polarization in
\texttt{supports} stays generally high in the whole period of analysis but
after the election, which suggests that some election winners might
concentrate support, lowering party polarization along coalition structures.
The patterns of polarization changes become evident when comparing the amount
of intra- and inter-party connections in different periods:  In the months of
September and October 2011 the polarization before the election manifests in
618 (54.4\%) \texttt{likes} within parties and 518 (45.6\%) \texttt{likes}
across parties, in comparison to the low polarization period a year before, in
September and October 2010, when there were 300 (33.7\%) \texttt{likes} within
parties and 591 (66.3\%) \texttt{likes} across parties. These observations,
while sensitive to the size and activity in the network, appear in contrast
with the control scenario of \texttt{comments}, in which no artificial pattern
of party polarization appears in the whole period.



\section{Party structures} 

In this section, we focus in two aspects of online political activity that are
not captured by polarization metrics, i.e. intra-party structures and inter-
party connectivity. In particular, we want to answer two questions: Do parties
with different ideologies create different social structures in  online
communities? And do parties with similar positions in political space connect
more to each other, despite the general pattern of polarization?

\subsection{Intra-party structures}

Previous research on the online political activity of users in the US
discovered  that right-leaning users showed higher online social cohesion than
left-leaning users \citep{Conover2012}. In the following, we extend the
quantification of each party both in its social structure and position in
ideological space. With respect to the latter, we locate the ideology of each
party in the two-dimensional space of  Left-Right and Conservative-Liberal
dimensions. To ensure the statistical relevance of our metrics, we restrict
our analysis to the 6 parties with more than $200$ politicians each, which cover
the two- dimensional spectrum of Swiss politics.

We analyze the social structure of each party based on the \texttt{supports}
subnetwork among the politicians of the party, capturing the asymmetric
relationships that lead to prestige and popularity. On each of these party subnetworks, we quantify three metrics related to
relevant properties of online social networks: hierarchical structures through
in-degree centralization, information efficiency through average path length,
and social resilience through maximum k-core numbers (see Appendix D). While these three metrics are not independent from
each other, they capture three components of online political activity that 
potentially differentiate parties: how popular their leaders are in
comparison to a more egalitarian structure, how efficient their social
structure is for transmitting information, and how big is the core of densely
connected politicians who would support each other under adverse conditions.

\begin{figure}[h]
\centering
\subfloat{\includegraphics[width=\textwidth]{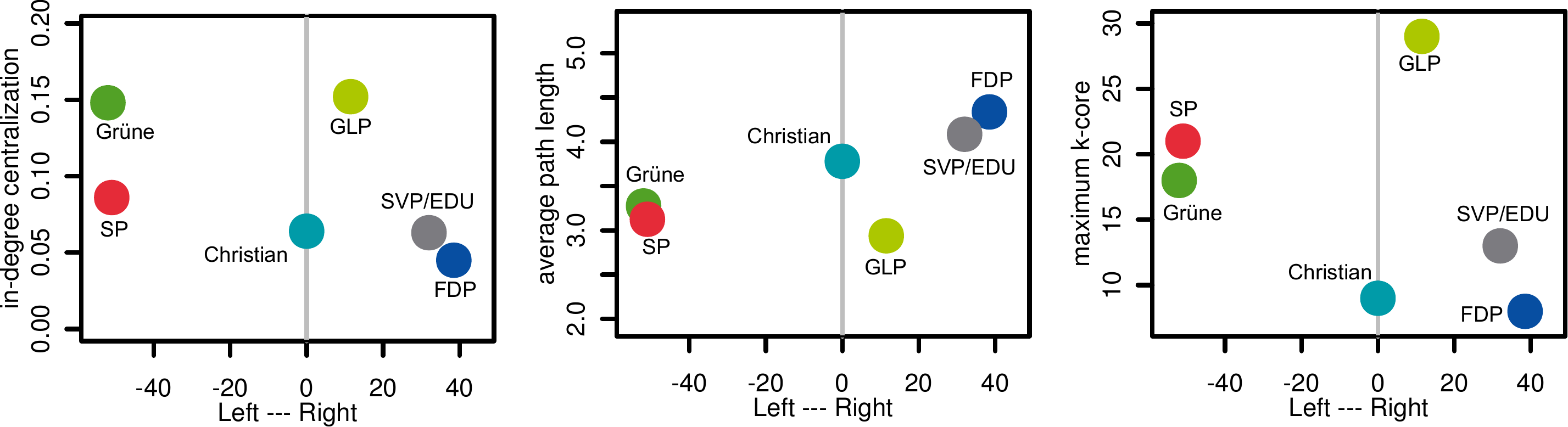}}\\
\subfloat{\includegraphics[width=\textwidth]{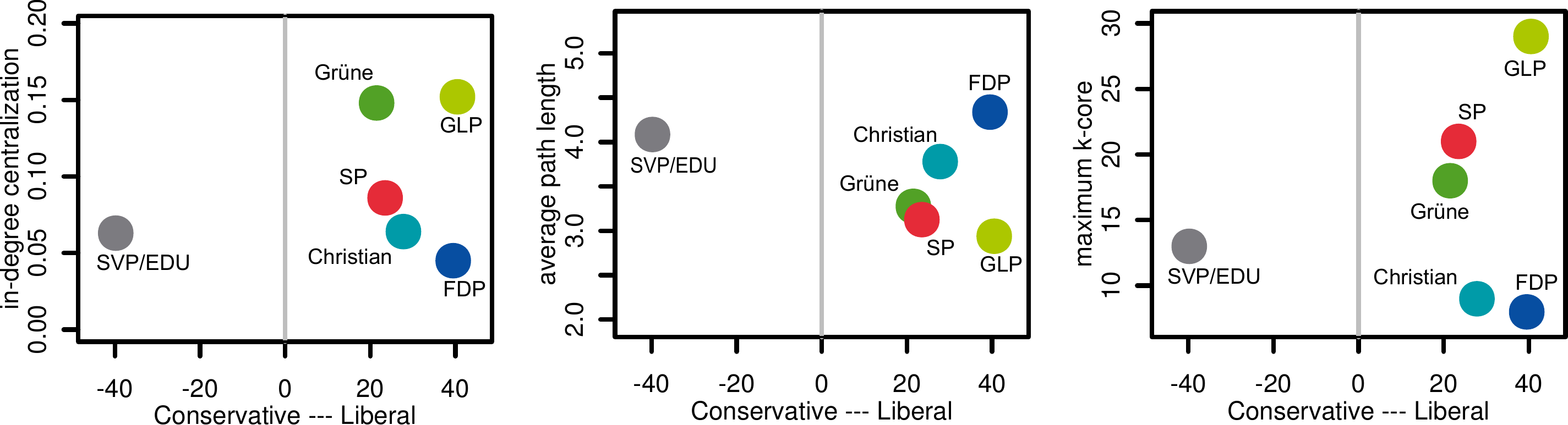}}
\caption{Social network metrics versus  their position in ideological space,
\textit{Left-Right} position (top), \textit {Conservative-Liberal} position
(bottom). The \texttt{supports} subnetworks of parties with federal
representation and more than 200 politicians each in \texttt{politnetz.ch}.}
\label{fig:supportsIntra} 
\end{figure}

Figure \ref{fig:supportsIntra} shows the value of the three social network
metrics versus the position of each party in both dimensions. The two parties
in the farthest right part of the spectrum, SVP and FDP, show higher average
path lengths and lower maximum k-core numbers than left-wing parties such as
SP and Gr\"une. This points to a difference in online activity in Switzerland
dependent on the the political position of parties: right parties have created
online social networks with lower information efficiency and lower social
resilience. This poses a contrast with previous findings for US politics, in
which politically aligned communities displayed the opposite pattern. This
leads to the conclusion that the position of an online community in the left-
right political spectrum does not universally define the properties of its
social structure, and that particularities of each political system create
different patterns.

There is no clear pattern of in-degree centralization in the \textit{Left-Right} 
 or \textit{Liberal-Conservative} dimensions,  but green parties (GLP
and Gr\"une) show a significantly higher in-degree centralization than the
rest. This result can be explained by the structure of the Swiss government,
which is composed of seven politicians from a coalition of various parties. At
the time of the study, this coalition includes politicians from the other four
parties, excluding both GLP and Gr\"une. This would give an incentive to these
parties to highlight a relevant member in order to gain a seat in the seven-
member government, and thus creating higher in-degree centralization in online
media.


\subsection{Inter-party connectivity}

Our second question tackles the activity across parties with respect to their
political position, hypothesizing that parties closer in ideological space will
be more likely to connect to each other. To do so, we first require a measure
to estimate the tendency of one party to connect to another under the presence
of polarization. To complement our analysis of Q-modularity, we use
demodularity between pairs of parties, comparing their tendency to connect
with what could be expected from a random network (see Appendix E).  Negative demodularity  indicates that a
party consistently avoids interaction with another party, contrary to positive
demodularity, which indicates that interactions are more frequent than expected at
random.

\begin{figure}[ht]
\centering
\includegraphics[width=0.33\textwidth]{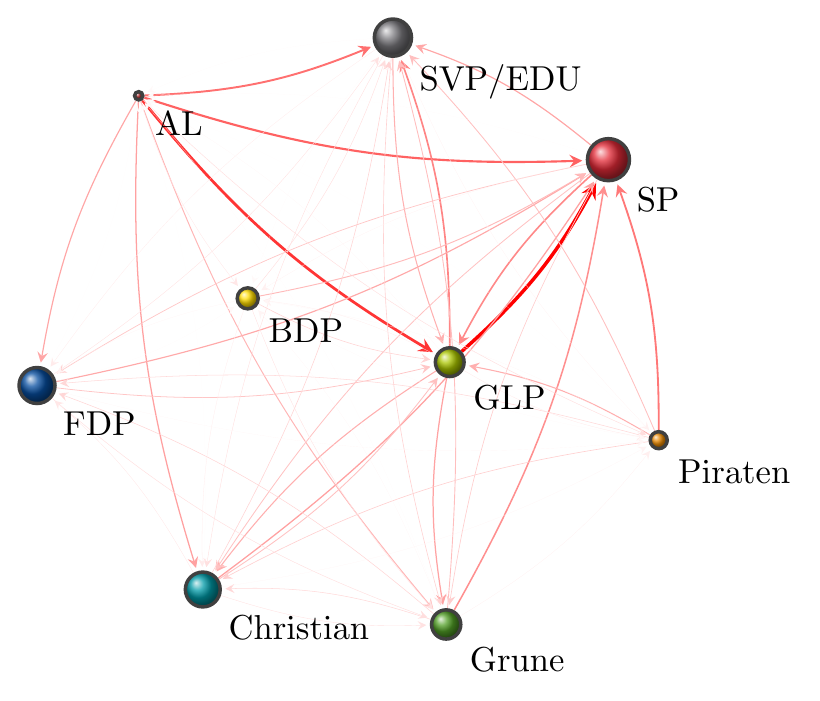}  
\includegraphics[width=0.32\textwidth]{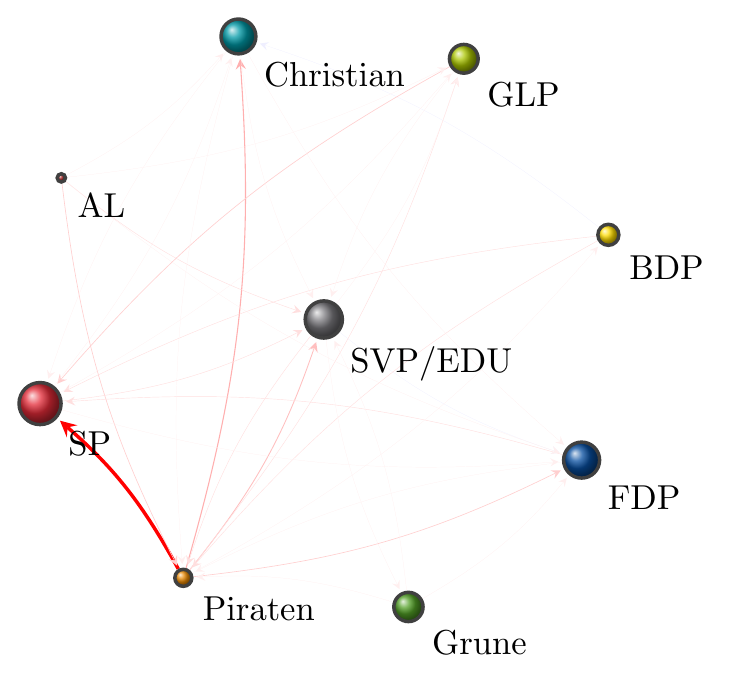}  
\includegraphics[width=0.33\textwidth]{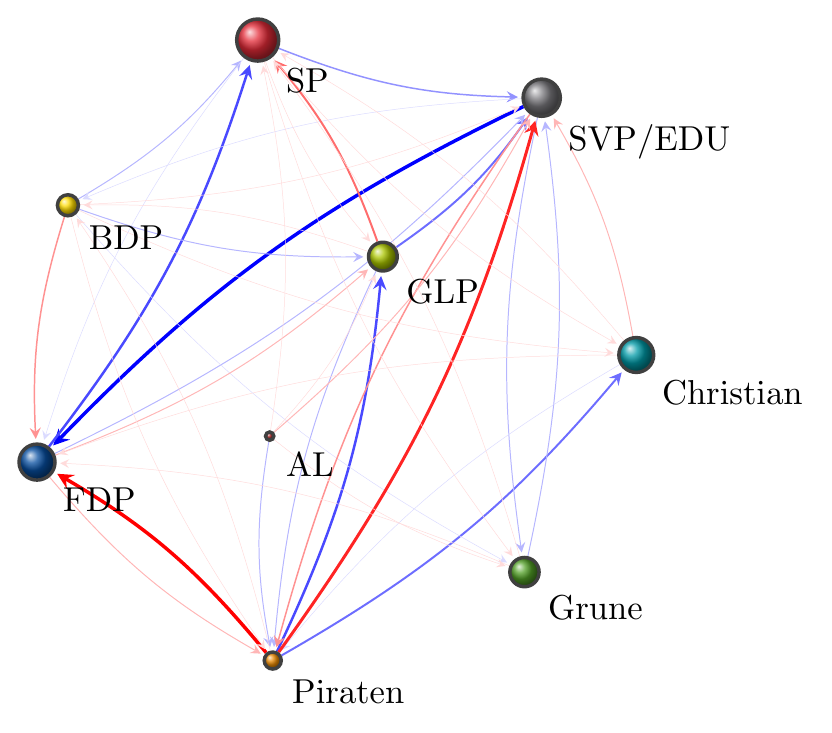}  
\caption{Networks of demodularity across parties for
  \texttt{supports} (left), \texttt{likes} (centre), and
  \texttt{comments} (right). Link width and color intensity is
  proportional to demodularity score between  parties, colored blue
  for positive values and red for negative
  ones. Node size is proportional to the amount of politicians of each 
party. Demodularity scores are strongly negative for supports, weakly 
negative for likes, and both positive and negative for comments.} 
\label{fig:demodNet}
\end{figure}

Figure \ref{fig:demodNet} shows a visualization of the network of politicians
aggregated by parties, for each layer of interaction. It can be observed that
demodularity scores have strong negative values in the \texttt{supports}
layer,  where all links have a negative weight (more details in Appendix E). This is in line with the strong
polarization among party lines, which makes \texttt{support} links to stay within
parties. In addition, the overall level of outgoing negative demodularity
shows a certain level of heterogeneity, as some parties have a stronger tendency
to not support any politician from another party. In the \texttt{likes} layer,
demodularity scores are  less negative, indicating that there are weaker
incentives to avoid giving a \texttt{like} to a politician of another party,
compared to \texttt{supports}. Positive scores are only present in the
\texttt{comments} layer, which shows that politicians are more likely to
comment posts of other parties, creating debates with opponents.

To relate the demodularity with the position in ideological space of parties,
we measured the Euclidean distance between pairs of parties in a two-dimensional
ideological space, explained in detail in Appendix F. Our hypothesis is that in layers of interaction
with positive connotation (\texttt{supports} and \texttt{likes}), parties that
are further from each other have weaker cross-party interactions, in contrast
to the \texttt{comments} layer, where links can be used to express disagreement
and thus political distance increases with cross-party interaction.  To test
these hypotheses, we compute the Pearson correlation coefficient of the
demodularity score of the pairs of parties versus the political Euclidean
distance between them (see Appendix G for details). This
correlation coefficient allows us to empirically test the existence of a
linear relationship between ideological distance and cross-party interaction
in each layer.

\begin{table}[h]
\centering
\begin{tabular}{|c|c|c|c|}
\hline
&Supports&Likes&Comments\\
\hline
$r_{\text{Pearson}}$ & -0.14  ($p=0.37$) & \textbf{-0.45} ($p<3\cdot10^{-3}$)   &  \textbf{0.45} ($p<2.7\cdot10^{-3}$) \\
\hline
\end{tabular}
\caption{Pearson correlation coefficients of the demodularity scores between parties
  and their pairwise Euclidean distance. }
\label{tab:demodEucl}
\end{table}

Table \ref{tab:demodEucl} shows the results for the three correlation
coefficients. First, the \texttt{supports} layer shows no significant
correlation, as data is not sufficient to reject the null hypothesis that both
variables are not related. This points to the high level of polarization in
\texttt{supports}, which makes links across parties so scarce that
demodularity scores are equally negative for parties close and far in
ideological space. The layer of \texttt{likes} shows a significant negative
correlation, indicating that demodularity is higher between parties with
closer ideologies. On the other hand, the correlation in the \texttt{comments}
layer is positive,  showing that politicians are more likely to participate in
online debates with politicians of parties that hold opposite views. This
highlights the meaning of \texttt{comments}, which are mainly used to argue
with politicians of other parties as opposed to \texttt{likes}, which are used
to agree with politicians with similar views, even though they might not
belong to the same party.

\section{Discussion}

Our work explores behavioral aspects of political polarization,
measuring network polarization over the digital traces left by
politicians in \texttt{politnetz.ch}. Our approach is centered around
the construction of a multiplex network with politicians as nodes and
three layers of directed links: one with \texttt{support} links, a second one
with link weights as the amount of \texttt{comments} a politician made to
another politician, and a third one with weights counting the amount
of times a politician \texttt{liked} the posts of another.  We studied network
polarization in the three layers as the level of intra-party cohesion
with respect to inter-party cohesion, measuring network polarization
as the modularity with respect to party labels, using Newman's
Q-modularity metric. This methodology allows us to investigate
polarization at a scale and resolution not achievable by traditional
opinion survey methods, including the time evolution of polarization
on a daily basis.  Furthermore, we provide a quantitative analysis of
the ways in which politicians utilize participatory media, the content
of discussion groups between politicians of different parties, and the
conditions that increase and decrease online network polarization.

We compared the information shared across the three layers, and found that
each layer contains a significant amount of link and community information
that is not included in any other layer.  The layers  of \texttt{supports} and
\texttt{likes} revealed significant patterns of polarization with respect to
party alignment, unlike the \texttt{comments} layer, which has negligible polarization.
This is particularly interesting with 
respect to opinion dynamics models, which frequently assume that the presence of 
opinion polarization implies a bias in the underlying communication network. While polarization is 
clearly present on politnetz with respect to like and support links, it does not seem to have decisive
influence on overal communication. 
We applied community detection algorithms at all three layers, and compared
the computationally found groups with the parties of the Swiss system.  At the
\texttt{comments} layer,  the community detection algorithms reveal that a partitioning
of politicians conveys higher modularity than party alignment, suggesting that
groups in \texttt{comments} are not party-driven. This is confirmed by a content analysis
of the \texttt{comments} in each group, where the most informative terms show that
each group discusses different topics. At the \texttt{supports} layer, the
partition of politicians into parties is very similar to the maximal partition
found by algorithms, suggesting that party alignment is nearly the most
polarizing partition of politicians. 
Further analysis will be necessary to
test this observation, measuring how the party alignment of a politician might
be predicted by its social context. In addition, our work focuses on data from
politicians, constituting an analysis of elite polarization. Future works can
include datasets from the electorate at large, measuring mass polarization
from other digital traces such as blogs \citep{Adamic2005} and Twitter
\citep{Conover2011}.

We computed the time series of network polarization of each layer, revealing
how the polarization in \texttt{likes} increased significantly around the federal
elections of 2011, compared to moments without electoral campaigns.
Polarization in \texttt{supports} and \texttt{likes} showed a sharp decrease after the
elections, possibly as an effect of coalition-building processes, and polarization in 
\texttt{comments} was close to 0 for the whole study
period. The evolution of polarization in \texttt{likes} and \texttt{supports} reveals a relation
between online activity and political events, in which social interaction
becomes more influenced by party membership when elections are close. Our
approach to the time evolution of polarization can be applied to test the
influence of other political events, for example how referendums might
increase polarization along the opinions about the issue being voted.

The \texttt{comments} layer showed no party-alignment polarization, but the
computational detection of modular structures reveals a different partition of
politicians. Our analysis of the content of the \texttt{comments} in these
partitions reveals that they follow different topics, and that modularity in
\texttt{comments} can be attributed to similar interests and competences among
politicians. This is emphasized when analyzing demodularity across parties and
ideological distance: demodularity in \texttt{comments} increases with
ideological distance, showing that parties that are further from each other
tend to create debates in which politicians of different views
\texttt{comment} on each other's contributions. Furthermore, demodularity in
\texttt{likes} decreases with distance between parties, showing that the
missing polarization in \texttt{likes} is due to politicians acknowledging the
contributions of people from other parties with similar ideologies. It keeps
open to test the role of other possible aspects of the connectivity across
parties, including geographical distribution, party size, and year of
creation.  Further analysis can test if new parties connect to other new
parties across ideological distance and if their growths are correlated with
connectivity, similarly to the growth of new European parties with information
seeking \citep{Bright2013}.  At the level of individual politicians,  our
methods allow the analysis of the relation between individual reputation and
contribution to polarization, quantifying if strong biases in online behavior
are tied to intra-party leadership and election results.

We analyzed the social networks of politicians in \texttt{politnetz.ch} to
explore the relation between ideology and social structures in online
interaction. We found that  green parties (GLP and Gr\"une) have a higher in-
degree centralization than the rest, indicating that their internal structure
is more unequal with respect to popularity. Two possible explanations for this
are the current configuration of the Swiss government, which excludes these
two parties, or a hypothetical relation between environmental politics and
party organization. Left-aligned parties have lower average path length and
higher maximum k-core numbers than right-aligned ones, showing that left
parties create social networks with higher information spreading capabilities
than right parties. This result is in contrast with previous findings
\citep{Conover2012} which found the opposite pattern for the networks of US
Twitter users. In addition, demodularity metrics across parties indicate that
their connectivity in terms of \texttt{likes} increases with closeness in
political space, while the connectivity in terms of \texttt{comments}
increases with distance. These findings lead to a set of hypotheses to test in
other multi-party systems, in order to understand the conditions that link
online social network structures and the political position of parties. The
digital traces left by politicians in Canada \citep{Gruzd2014}, Germany
\citep{Lietz2014}, and Spain \citep{Aragon2013} are first potential candidates
for this kind of analyses.

Our results highlight that the strategies for campaigning and mobilization in
online media differ with respect to the political position of a party, and
that polarization in online behavior is heavily influenced by party membership
and upcoming elections.   Our findings show that the  previously found
relations between social network structures and ideology are not universal,
calling for new theories that apply for multi-party systems. We showed that
the analysis of network polarization through digital traces conveys an
alternative approach to traditional survey methods, capturing elements of the
time dependence and the phenomena that influence polarization. The multi-party
nature of Switzerland and the crowdsourced  online activity of its politicians
allowed us to test the relation between the ideological distance and online
interaction between parties, showing us a very clear picture of online
interaction.  First, \texttt{supports} are strongly biased towards politicians of the
same party. Second, \texttt{comments} happen across parties of different political
views, clustering politicians into groups with similar interests. And third,
\texttt{likes} cross party lines towards politicians of similar opinions, but this
effect is attenuated when elections are close, creating  a highly polarized
state that relaxes after elections are over.

\section{Acknowledgements} 
This work was funded by the Swiss National Science Foundation
(CR21I1\_146499). We thank politnetz.ch for their support on the data
retrieval.

\bibliographystyle{sg-bibstyle}
\bibliography{ipp}

\newpage

\appendix

\section{Appendix: Layer similarity metrics}
\label{sec:appendix}

\subsection{Jaccard Similarity Coefficient}
\label{sec:appendix:layersJaccard}

The Jaccard similarity coefficient is used to compare similarity and diversity between two
sets. Its value for sets $X$ and $Y$ is defined as a
division of the intersection of two sets over the union of these sets:
\begin{equation*}
J_{XY} = \frac{|X \cap Y|}{|X \cup Y|}
\end{equation*}

For computing one-sided overlaps, the \textit{Partial Jaccard coefficient}
normalizes over one set only. For instance, the $J_{XY}$ normalized over $Y$ gives
the fraction of the set $Y$ which is attributed to the set $X$: 
\begin{equation*}
\text{Partial }J_{XY} = \frac{|X \cap Y|}{|Y|} 
\end{equation*}

\subsection{Mutual information}
\label{sec:appendix:layersInfo}

\paragraph{Shannon's Entropy} 
Claude Shannon introduced in 1948 the following measure of
\textit{information} through the \textit{entropy} of the information
source. \textit{Entropy} stands for the measure of uncertainty of the
information content, or how much of the uncertainty will be reduced
when information is received. We illustrate a connection between the
words \textit{information} and \textit{entropy} on the example of a
coin toss. 

A biased coin with both sides having heads will always land heads as a result
of any toss event. If $X_{\text{biased}}$ is a random variable representing
the result of a coin toss, then $p(X_{\text{biased}}=\text{Heads}) = 1$ and
$p(X_{\text{biased}}=\text{Tails}) = 0$. Before the coin toss, our uncertainty
about the result is negligible, since we know that at every trial the coin
lands heads in anyway. We call that the entropy of such a process is small.
Furthermore, after tossing the coin we can also state that we learn nothing
new as outcome is known \textit{a priori}, that is the amount of the uncertainty
reduced is negligible, and thus the information received is also negligible.
Assume that we toss an unbiased coin with  equal probabilities of having heads
or tails, $p(X_{\text{unbiased}}=\text{Heads}) = \frac{1}{2}$ and
$p(X_{\text{unbiased}}=\text{Tails}) = \frac{1}{2}$. In this setup the
uncertainty we have about the result of the event is the highest, since this
coin can land heads or tails with equal probabilities. The entropy of
this process is high, and thus the amount of the information received after each coin
toss is also high.

If the result of a random variable is difficult to predict, then the
entropy, unpredictability or uncertainty, of the outcome is high; conversely, if the result
is predictable, it's entropy is low. Entropy is at its highest value,
when the outcomes of a random variable are equally probable. \citet{Shannon1948} proved
that only a logarithmic function satisfies the properties necessary to
measure the entropy, and proposed the following formula:
\begin{equation*} 
H(A) = -\sum \limits_{i=1}^{m} p(a_i) \cdot log_b p(a_i)
\end{equation*} 
where $A$ is a discrete random variable taking $m$ possible
values   $(a_1, a_2, \dots, a_m)$, $p(a_i)$ is  the probability of $A$ taking
value $a_i$, and $b$ is the base of the logarithm used which
determines the information units of the entropy. For $b=2$, the
entropy is measured in \textit{the number of bits per random variable
  outcome}. Additionally, \textit{information} can be expressed in
terms of the entropy as: 
\begin{equation*}
I = k \cdot H(A)
\end{equation*}
where $k$ is the number of events. Therefore, with the 
logarithm of base $2$, information is measured in \textit{bits}.

Thus, a coin that always falls heads has the entropy of $H(X_{\text{biased}}) =
-(0 \cdot log_2 0 + 1 \cdot log_2 1) = 0$ bits per outcome which gives us no
information after each throw, i.e.  $I = 1 \cdot 0 = 0$ bits. On the other
hand, the entropy of an unbiased coin is $H(X_{\text{unbiased}}) =
-(\frac{1}{2} \cdot log_2 \frac{1}{2} + \frac{1}{2} \cdot log_2 \frac{1}{2}) =
1$ bit per outcome, and the information received is $I = 1 \cdot 1 = 1$ bit.

\paragraph{Miller-Madow Entropy Estimator} Calculation of the entropy
from the data samples requires the knowledge on the probabilities of
the outcome of the random variable as seen from the Shannon's
formula. In practice, $p(a_i)$ is unknown and must be
\textit{estimated} from the observed counts of the $i$-th outcome in
the sequence of trials. The simplest and widely used estimator of the
entropy is the \textit{Maximum Likelihood estimator (ML)},
\citep{Hausser2009}:
\begin{eqnarray*}
\hat{H}^{ML}(A) &=& -\sum \limits_{i=1}^{m} \hat{p}^{ML}(a_i) \cdot log_2
\hat{p}^{ML}(a_i) \\
\end{eqnarray*}

where $\hat{p}^{ML}(a_i)$ is an estimate of the probability of the
$i$-th possible outcome $a_i$ of the random variable, often calculated
as $\hat{p}^{ML}(a_i) = \frac{y_i}{n}$, where $y_i \geq 0$ is the
number of the observed counts of the $i$-th outcome out of the $n$
number of observations. 

When the number of the possible outcomes of the random variable is
much lower than the number of observations, $m \ll n$, the ML method
gives the optimal estimation of the entropy. On the other hand, the finite
sample sizes can leave some outcomes unobserved, and the ML method can
underestimate the true entropy. To overcome the sample size bias,
corrections and other estimators of the entropy have been developed
\citep{Hausser2009}. For our empirical estimation of the information
content, we employ the  \textit{Miller-Madow entropy estimator}
\citep{Miller1955}. In essence, it is the ML entropy with the bias
correction: 
\begin{equation*} 
\hat{H}^{MM}(A) = \hat{H}^{ML}(A) + \frac{p_{>0}-1}{2 \cdot n} 
\end{equation*} 
where $p_{>0}$ is the number of outcomes with $y_i > 0, \forall i \in [1..m]$,
i.e. how many of the $m$ possible outcomes were observed in $n$ trials.

\paragraph{Mutual Information} If $X$ and $Y$ are two random
variables, then the mutual information between them is expressed
as:
\begin{equation*}  I(X;Y) = 
H(X) - H(X|Y) = H(X) + H(Y) - H(X,Y) \end{equation*}
where H(X), H(Y) are the marginal entropies, or simply the entropies
of $X$ and $Y$; $H(X,Y)$ is the joint entropy of $X$ and $Y$, and
$H(X|Y)$ is the conditional entropy of $X$ given $Y$, which is the
amount of the uncertainty that remains in $X$ when the value of $Y$ is
known. The mutual information measures the average reduction of the
uncertainty in $X$ that results from learning $Y$; and vice versa
\citep{MacKay2002}. 

\begin{figure}[!h]
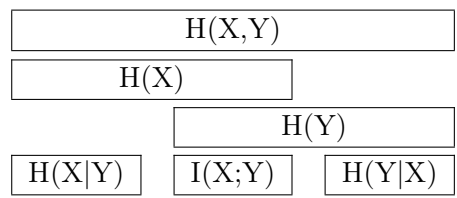

\centering
\begin{tabular}{cccccc}
\hline
\multicolumn{6}{|c|}{H(X,Y)}\\ \hline \\[-2.5ex] \cline{1-4}
\multicolumn{4}{|c|}{H(X)}&&\\\cline{1-4} \\[-2.5ex]  \cline{4-6}
&&&\multicolumn{3}{|c|}{H(Y)}\\\cline{4-6}\\[-2.5ex]  \cline{1-2} \cline{4-4} \cline{6-6}
\multicolumn{2}{|c|}{H(X|Y)}&&\multicolumn{1}{|c|}{I(X;Y)}&&\multicolumn{1}{|c|}{H(Y|X)}\\
\cline{1-2} \cline{4-4} \cline{6-6}
\end{tabular}
\caption{The relationship between the joint entropy, the marginal
  entropy, the conditional entropy and the mutual information
  \citep{MacKay2002}.} 
\label{fig:entMI}
\end{figure}

The mutual information is symmetric, $I(X;Y) = I(Y;X)$, and is
depicted in relation to the entropy in Figure \ref{fig:entMI}.

The mutual information can be measured in terms of the
probabilities of the outcomes of the random variables:
\begin{equation*}
I(X;Y) = \sum \limits_{y \in Y} \sum \limits_{x \in X} p(x,y)\cdot log_2
\left ( \frac{p(x,y)}{p(x) \cdot p(y)} \right )
\end{equation*}
where $p(x,y)$  is the joint probability distribution of $X$ and $Y$, and
$p(x)$  and $p(y)$ are the marginal probabilities of the outcomes $x$ and $y$. We
compute the entropy and the mutual information using the
\textbf{\textsf{Entropy}} library of the \textbf{\textsf{R}}
programming language developed by \citet{Hausser2009}.

\paragraph{Normalized Mutual Information} This metric is also known as
the \textit{uncertainty coefficient}, \citep{Press2007}. The uncertainty
coefficient of $X$ with respect to $Y$ is given by:
\begin{eqnarray*}
\text{NMI}(X|Y) = \frac{I(X;Y)}{H(X)} = \frac{H(X) - H(X|Y)}{H(X)}
\end{eqnarray*}

In essence, the uncertainty coefficient of the dependent variable $X$ with
respect to the independent variable $Y$ is the mutual information of both
variables normalized over the entropy of the dependent variable. The
measure lies between $0$ and $1$, and is interpreted as follows: if
the NMI$(X|Y) = 0$, then there is no relation between $X$ and $Y$; if
the NMI$(X|Y) = 1$, then the knowledge of $Y$ fully predicts or
determines $X$, or 100\% of the information content of $X$ is captured
by $Y$. A value in-between $0$ and $1$ gives the fraction of the
information gain of $X$ when $Y$ is known. Interchanging $X$ and $Y$
in the uncertainty coefficient, NMI$(Y|X)$, will define the dependence
of $Y$ with respect to $X$, and the normalization is performed over
the entropy of $Y$.

\section{Appendix: Measuring network polarization}

\subsection{Q-modularity} 
\label{sec:appendix:modularity}  

In the network theory, a network shows a modular structure if it can be
partitioned into the groups that are densely interconnected inside and loosely
connected to the other groups. The \textit{Q-modularity} metric quantifies the
quality of a partition in terms of a modular structure \citep{Newman2006},
comparing the fraction of links of nodes within the groups to the expected value
if links were distributed purely at random, but preserving the nodes' degrees. The formula for
a \textit{directed} network is given by:
\begin{equation*}
Q = \frac{1}{m} \cdot \sum\limits_{i=1}^N\sum\limits_{j=1}^N \left [ A_{ij} - \frac{k_i^{\text{out}}\cdot k_j^{\text{in}}}{m} \right ] \delta(i,j)
\end{equation*}
where $N$ and $m$ are the amounts of nodes and links in the network,
$k_i^{\text{out}}$ and $k_i^{\text{in}}$ are the out-degree and the in-degree of
node $i$, $A$ is the adjacency matrix of the network,  and $\delta(i,j)$ is a
function that takes the value 1 if nodes $i$ and $j$ are in the same
group, and $0$ otherwise. For the case of a weighted network, the amount
of links is replaced with the sum of weights of all links, and the adjacency
matrix has entries corresponding to the weight of each link.

\subsection{Community detection algorithms}
\label{sec:appendix:algorithms}  

In an arbitrary network, finding a partition with the maximum Q-modularity is
known as the \textit{community detection problem}. There is a wide variety of
algorithms producing different network partitions which aim at finding
a partition with the maximum Q-modularity. To find partitions with the
maximal modularity in each layer, we used the state-of-the-art
community detection algorithms from the open source software
\textbf{\textsf{Radatools}}, \citep{Radatools}. Following is the list
of the algorithms that are implemented in the software, abbreviations
of the algorithms are given in brackets:

 \begin{itemize}
 \item (e): extremal optimization, see \citep{Duch2005}
 \item (s): spectral optimization, see \citep{Newman2006}
 \item (f): fast algorithm, see \citep{Newman2003}
 \item (r): fine-tuning by reposition
 \end{itemize}

The algorithms (s) and (e) have a stochastic behaviour, thus they must be executed
with several repetitions, e.g. (esrfr-30) means perform 30 repetitions
of the algorithm (e), 30 repetitions of (s), 1 times (r), 1 times (f) and finally 1
times (r), refer to the \citep{Radatools} for more details.  We run  the
following combinations of the algorithms for each layer: e-1, esrfr-30, r-1, f-1,
s-10, rfr-1, rsrfr-30, and store the partition that gives the highest
Q-modularity from among all the runs. This optimal partition, which is computed
independently from the party alignment, determines the  $Q_{comp}$ polarization
score. We repeated this analysis for each layer, which resulted in
three values of the $Q_{comp}$: one for \texttt{supports}, one for
\texttt{likes}, and another one for \texttt{comments}.

\subsection{Time series of network polarization}
\label{sec:appendix:timeseries}  

\begin{figure}[ht]
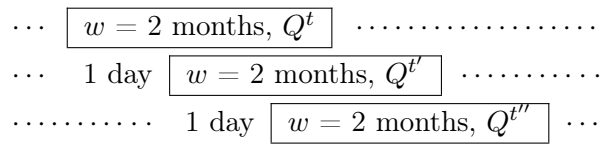

\centering
\begin{tabular}{ccccccc}
 \cline{2-4}
$\cdots$&\multicolumn{3}{|c|}{$w$ = 2 months, $Q^t$}&\multicolumn{3}{c}{$\cdots\cdots\cdots\cdots\cdots\cdots\cdot$}\\ 
\cline{2-4} \\[-2.5ex] \cline{3-5}
$\cdots$&\multicolumn{1}{c}{1 day}&\multicolumn{3}{|c|}{$w$ = 2 months, $Q^{t^{\prime}}$}&\multicolumn{2}{c}{$\cdots\cdots\cdots\cdot\cdot$}\\
\cline{3-5} \\[-2.5ex]  \cline{4-6}
\multicolumn{2}{c}{$\cdots\cdots\cdots\cdot\cdot$}&\multicolumn{1}{c}{1 day}&\multicolumn{3}{|c|}{$w$ = 2 months,
  $Q^{t^{\prime \prime}}$}&$\cdots$\\
\cline{4-6}\\
\end{tabular}
\caption{Time-series segmentation approach of computing modularity
  score of a network with a sliding window. At each shift of the
  window, modularity is computed on the network with the links
  present within the current time window. The size of the window is set at
  2 months, the step of the window shift is 1 day.}
\label{fig:time}
\end{figure}

We construct a time series of polarization with a \textit{rolling window}
technique, dividing the network data along the time axis in the overlapping
intervals of a fixed size. Each interval overlaps with the next one on a fixed
subinterval, where the difference in their starting dates is a constant
\textit{step} size. This technique smooths out the data through the
aggregation of links within the window, and preserves a granularity defined by
the step size. For our analysis, we chose a window of 2 months and a 1-day
step, in order to aggregate sufficient data in each window and to preserve daily
resolution in the time series. In every time interval, we compute the
Q-modularity on the network with links that have timestamps within the
specified interval. We illustrate the method with an example and Figure
\ref{fig:time}. At the timestamp of the first link, $t$, we obtain the first-
window modularity score for the interval $[t,t +\text{2 months}]$, then we
move the window one step forward to the starting time $t' = t + \text{1 day}$,
recompute the metric on the network with the links present within the new
window, to obtain the modularity score at the time $t'$. We continue sliding the
time window, computing the modularity until reaching the timestamp of the last
link. Hence, with this approach, we record the evolution of the
network polarization around the dates in each time window.

\section{Appendix: Word information content}
\label{sec:appendix:PMI}

\paragraph{Pointwise Mutual Information} The PMI is the mutual information for
\textit{the pairs of outcomes} of two random variables rather than all
possible values. Thus, there is a direct link between the MI and the 
PMI: the mutual information of the two random variables $X$ and $Y$ is the
expected value of the PMI over all possible pairs of outcomes of the
two random variables. 

\begin{equation*}
\text{PMI}(X=x;Y=y) = log_2 \frac{p(x,y)}{p(x) \cdot p(y)} =  log_2
\frac{p(x|y)}{p(x)} = log_2 \frac{p(y|x)}{p(y)}
\end{equation*}
where $p(x,y)$ is the joint probability distribution of the outcomes $x$
and $y$ of the random variables $X$ and $Y$ respectively, and $p(x)$ and $p(y)$
are the marginal probability distributions of the outcomes $x$ and $y$ of the
random variables $X$ and $Y$ respectively.

We use the PMI to detect the information that a word contains within a set of
comments in a  group of the \texttt{comments} layer, approximating $p(x)$ as the
total frequency of the word in all comments, and $p(x|y)$ as the frequency of
the word within the comments of the group. To control for statistical
significance, we performed $\chi^2$ tests on the ratio of frequencies,
filtering out words with the PMI significance below the 99\% confidence level.
This gives us a list of words with significantly high PMI for each comment
group, indicating the most informative terms of the discussions within each
group of politicians.

\section{Appendix: Intra-party social network metrics}
\label{sec:appendix:sna}

\subsection{In-degree Centralization} 
\label{sec:appendix:sna-centralization}
\citet{Freeman1978} introduced the  concept of degree centralization, where
the average difference between the node degree centralities is normalized over the
value for a star network. In our case, we use this idea to capture
hierarchical structure inside the parties through the in-degree centralization:
\begin{equation*}
C_{in} = \frac{\sum\limits_{i=1}^n \left [ k^{in}_* - k^{in}_i\right
  ]}{\text{max} \sum\limits_{i=1}^n \left [ k^{in}_* - k^{in}_i \right ]}
\end{equation*} 
where  $k^{in}_i$ is the in-degree of node $i$, $k^{in}_*$ is the
largest in-degree of the network and ${\text{max} \sum\limits_{i=1}^n
  \left [ k^{in}_* - k^{in}_i \right ]}$ is the maximum possible sum
of differences in the degree centrality, which corresponds to the
value of a star network. 

The numerator represents the sum of differences between the highest degree in
the graph given by node $v*$ and the degrees of the other nodes, measuring the
extent to which the most central node $v*$ exceeds the in-degree of
the other nodes. The denominator stands for the maximum possible value of such
difference in the network with the same number of nodes. \citep{Freeman1978}
showed that the maximum difference is achieved in the networks of a  star-like
structure where one node dominates the network with respect to its centrality,
and produces the degree centrality of the highest value of $1$; conversely,
the lowest boundary of $0$ is obtained in complete networks, e.g. cliques,
where all possible links between nodes are present, and thus, nodes are
homogeneous and equal. Normalized over the denominator, the degree centrality is a
value in the interval $[0,1]$ and represents the average deviation of nodes in
the network from the most central node.

\subsection{Average Path Length} 

\label{sec:appendix:sna-path}
This social network metric measures the efficiency of the information
transportation in a network:
\begin{equation*}
l(G) = \frac{1}{n \cdot (n-1)} \cdot \sum\limits_{i=1}^{n}\sum\limits_{j=1}^{n} d(v_i,v_j),
\forall i\neq j
\end{equation*}
where  $d(v_i,v_j)$ is the length of the shortest path between nodes $v_i$ and
$v_j$, and $n$ is the number of nodes in the network. In essence, the
average path length is the sum of the lengths of the shortest paths between
all pairs of nodes in the network divided over the maximum number of
all possible pairs, $\forall i,j$ s.t. $i \neq j$, in the network
where a path can exist, therefore the normaliztion factor is $\frac{1}{n
\cdot (n-1)}$. If the network consists of the disconnected components,
the normalization factor becomes the number of existing paths in the
network: 
\begin{equation*}
l(G) = \frac{1}{\sum\limits_{i=1}^{n}\sum\limits_{j=1}^{n}\delta(i,j)} \cdot \sum\limits_{i=1}^{n}\sum\limits_{j=1}^{n} d(v_i,v_j),
\forall i\neq j
\end{equation*}
where $\delta(i,j)=1$ if there exists a path between the nodes $i$ and
$j$, $0$ -- otherwise.

\subsection{\textit{k}-core Decomposition} 
\label{sec:appendix:sna-kcore}

\begin{figure}[h]
  \centering
  \includegraphics[width=0.8\textwidth]{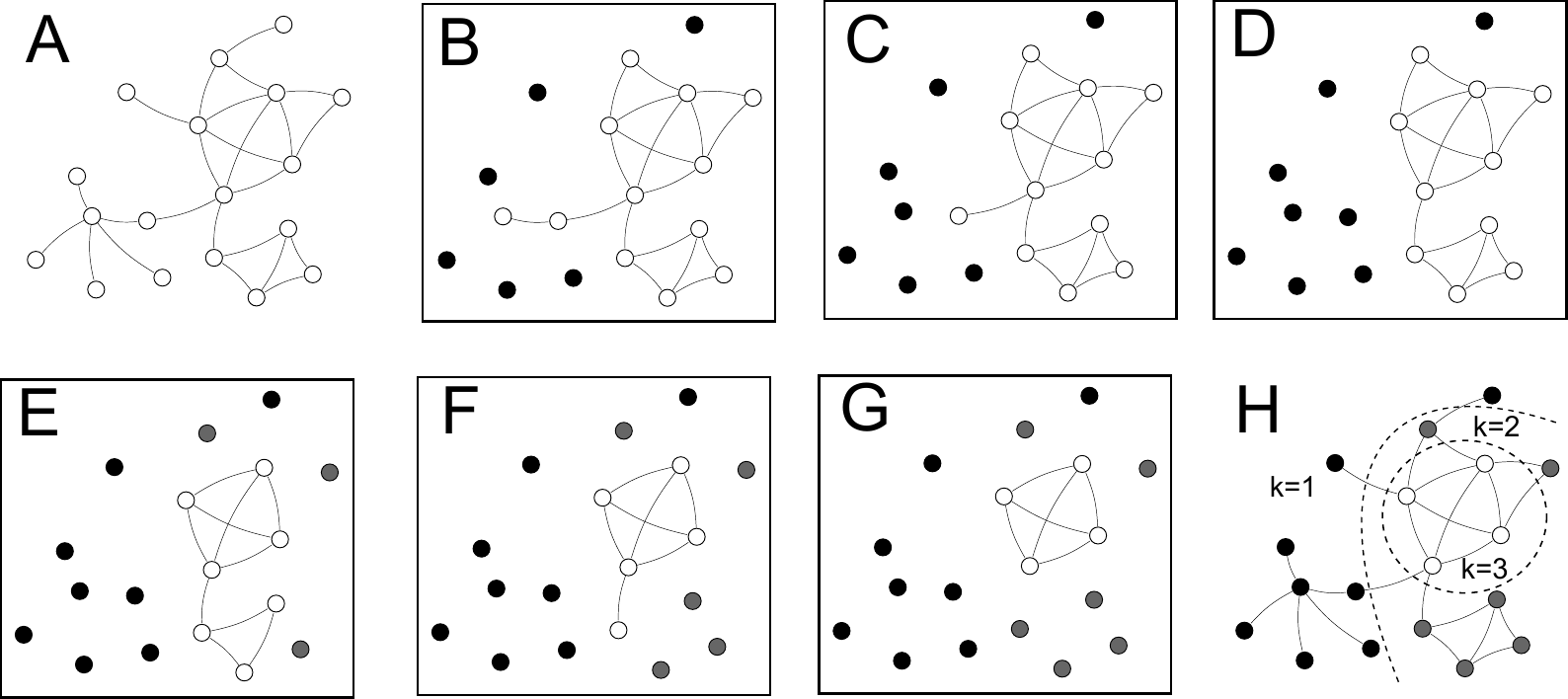}
  \caption{Effects of node removals on network connectivity as captured
    by degree only (A $\rightarrow$ B) and $k$-core decomposition (A
    $\rightarrow$ C $\rightarrow$ D $\rightarrow$ E)} 
  \label{fig:schema}
\end{figure}

A $k$-core of a network is a sub-network in which all nodes have a degree
$\geq k$. The $k$-core decomposition is a procedure of finding all $k$-cores,
$\forall k > 0$, by repeatedly pruning nodes with degrees $k$. Therefore, it
captures not only the direct, but also the indirect impact of users leaving
the network. As an illustration consider Figure \ref{fig:schema}, which shows
the process of finding the $k$-core decomposition:

\label{sec:resilience}

Starting again from A, and applying the $k$-core procedure, will repeatedly
remove nodes of degree less than 2, until only those with degree of at least 2
remain in panel D. The removed nodes up to this point are in the 1-core of the
network but not in the 2-core. In a second iteration, all nodes of degree less
than 3 are removed,  colored dark gray in panel G. These form part of the 2-core of
the network, but not of the 3-core, which is composed of the last four nodes.
Hence, supposing that users leave a community when they are left with less
than 3 friends, the $k$-core decomposition captures the full cascading effect
that departing users have on the network as a whole. More details on the
empirical calculation of the k-core decomposition can be found elsewhere
\citep{Garcia2013}.

\section{Appendix: Demodularity} 
\label{sec:appendix:de-modularity}  

We introduce an inter-group measure called \textit{demodularity},
which quantifies the relationship across different groups rather than
within the group, and can be defined as a property of a network where
nodes of one group preferentially attach to the nodes of the other
group. Negative scores of the demodularity from group $f$ to group $t$
indicate that nodes of $f$ strongly avoid interactions with nodes of $t$,
contrary to positive scores that show cross- community
interactions. The demodularity, $\overline{Q}_{ft}$, from community $f$ to
community $t$ in a \textit{directed} network is defined as:
\begin{equation*}
\overline{Q}_{ft} = \frac{1}{m_f} \cdot \sum\limits_{i=1}^N\sum\limits_{j=1}^N \left [ A_{ij} - \frac{k_i^{\text{out}}\cdot k_j^{\text{in}}}{m} \right ] \delta(C(i),f) \cdot \delta(C(j),t)
\end{equation*}
where $C(i)$ is the group to which node $i$ belongs, $\delta(C(i),f)$ is a
function such that $\delta(C(i),f) = 1$   if the group of node $i$ is $f$, and
$0$  otherwise, and the rest of the notation is consistent with the definition
of Q-modularity of Appendix \ref{sec:appendix:modularity}.

Tables \ref{tab:demodSup}, \ref{tab:demodLik}, \ref{tab:demodCom} show the
demodularity scores for the layers of \texttt{supports}, \texttt{likes} and
\texttt{comments}.

\begin{table}[b]
\begin{tabular}{c|c|c|c|c|c|c|c|c|c}
&AL&BDP&Christian&FDP&GLP&Grune&Piraten&SP&SVP/EDU\\
\hline
AL&-&-0.012&-0.046&-0.043&\cellcolor{red!25}-0.099&-0.036&-0.016&\cellcolor{red!25}-0.077&\cellcolor{red!25}-0.07\\
BDP&-0.002&-&-0.01&-0.006&-0.021&-0.014&-0.004&-0.034&-0.016\\
Christian&-0.002&-0.003&-&-0.012&-0.031&-0.014&-0.006&-0.046&-0.021\\
FDP&-0.002&-0.003&-0.013&-&-0.028&-0.016&-0.005&-0.042&-0.014\\
GLP&-0.005&-0.009&-0.039&-0.033&-&-0.045&-0.011&\cellcolor{red!50}-0.125&\cellcolor{red!25}-0.06\\
Grune&-0.003&-0.005&-0.019&-0.02&-0.044&-&-0.008&\cellcolor{red!25}-0.055&-0.032\\
Piraten&-0.002&-0.005&-0.023&-0.019&-0.039&-0.024&-&\cellcolor{red!25}-0.066&-0.031\\
SP&-0.003&-0.007&-0.03&-0.027&\cellcolor{red!25}-0.058&-0.024&-0.01&-&-0.043\\
SVP/EDU&-0.003&-0.003&-0.015&-0.01&-0.038&-0.022&-0.007&\cellcolor{red!25}-0.057&-\\
\end{tabular}
\caption{The demodularity scores of the layer of
  \texttt{supports}. Scores in the range $(-0.1,-0.05]$ are highlighted
in light red colour, and scores in the range $(-\infty,-0.1]$ are
displayed in red.}
\label{tab:demodSup}
\end{table}
\begin{table}[h]
\begin{tabular}{c|c|c|c|c|c|c|c|c|c}
&AL&BDP&Christian&FDP&GLP&Grune&Piraten&SP&SVP/EDU\\
\hline
AL&-&0&-0.001&-0.001&-0.001&0&-0.003&0&-0.002\\
BDP&0&-&0.001&0&0&0&-0.002&-0.002&0\\
Christian&0&0&-&-0.001&0&0&-0.001&-0.001&-0.001\\
FDP&0&0&0&-&0&0&-0.001&-0.002&0\\
GLP&0&0&0&-0.001&-&0&0&-0.003&-0.001\\
Grune&0&0&0&-0.001&0&-&-0.001&0&-0.001\\
Piraten&0&-0.001&-0.005&-0.003&-0.002&-0.003&-&-0.016&-0.004\\
SP&0&0&-0.001&-0.001&-0.001&0&-0.003&-&-0.002\\
SVP/EDU&0&0&0&0.001&-0.001&-0.001&-0.002&-0.003&-\\
\end{tabular}
\caption{The demodularity scores of the layer of \texttt{likes}.}
\label{tab:demodLik}
\end{table}
\begin{table}[h]
\begin{tabular}{c|c|c|c|c|c|c|c|c|c}
&AL&BDP&Christian&FDP&GLP&Grune&Piraten&SP&SVP/EDU\\
\hline
AL&-&0&0&0&-0.001&-0.001&\cellcolor{blue!25}0.002&-0.001&-0.002\\
BDP&0&-&-0.001&-0.003&\cellcolor{blue!25}0.002&\cellcolor{blue!25}0.001&-0.001&\cellcolor{blue!25}0.002&-0.001\\
Christian&0&0&-&-0.001&0&0&\cellcolor{blue!25}0.001&-0.001&-0.002\\
FDP&0&0&-0.002&-&-0.002&0&-0.002&\cellcolor{blue!50}0.005&\cellcolor{blue!25}0.002\\
GLP&0&-0.001&0&-0.005&-&0&\cellcolor{blue!25}0.002&-0.004&\cellcolor{blue!25}0.004\\
Grune&0&0&0&-0.001&0&-&0&-0.001&\cellcolor{blue!25}0.002\\
Piraten&0&-0.001&\cellcolor{blue!25}0.004&-0.007&\cellcolor{blue!50}0.005&-0.003&-&0&-0.006\\
SP&0&0&-0.001&\cellcolor{blue!25}0.001&-0.001&-0.001&0&-&\cellcolor{blue!25}0.003\\
SVP/EDU&0&\cellcolor{blue!25}0.001&0&\cellcolor{blue!50}0.007&0&\cellcolor{blue!25}0.002&-0.003&0&-\\
\end{tabular}
\caption{The demodularity scores of the layer of
  \texttt{comments}. Positive scores in the range $(0,0.005)$ are
  highlighted in light blue colour, and scores in the range
  $[0.005,\infty)$ are displayed in blue.}
\label{tab:demodCom}
\end{table}

\section{Appendix: Political distance} 
\label{sec:appendix:distance}  

The Euclidean distance between two points $\mathbf{X}$ and $\mathbf{Y}$ is the
length of  the line segment connecting them: $\mathbf{\overline{XY}}$. In
Cartesian coordinates, if $\mathbf{X} = (x_1, x_2, \dots, x_n)$ and
$\mathbf{Y} = (y_1, y_2, \dots, y_n)$ are two points in the Euclidean $n$-space,
then the distance between $\mathbf{X}$ and $\mathbf{Y}$ is defined as:
\begin{equation*}  \mathbf{d}(\mathbf{X},\mathbf{Y})=\sqrt{(x_1 - y_1)^2 +
(x_2 - y_2)^2  + \dots + (x_n - y_n)^2} \end{equation*}

Each of the parties $\mathbf{X}$ and $\mathbf{Y}$ has two political position
coordinates: \textit{Left-Right (lr)} and \textit{Liberal-Conservative (lc)}.
If $\mathbf{X} = (x_{lr},x_{lc})$  and $\mathbf{Y} = (y_{lr},y_{lc})$, then
the Euclidean party distance is given by:
\begin{equation*}
 \mathbf{d}(\mathbf{X},\mathbf{Y})=\sqrt{(x_{lr} - y_{lr})^2 + (x_{lc} - y_{lc})^2}
\end{equation*}

\section{Appendix: Correlation estimation}
\label{sec:appendix:corr}

The Pearson correlation coefficient is a measure of the association
between two variables $X$ and $Y$, and is defined as the covariance of
the two variables divided by the product of their standard deviation:
\begin{eqnarray*}
r = \frac{\sum\limits_{i=1}^n(x_i-\overline{X})\cdot(y_i-\overline{Y})}{\sqrt{\sum\limits_{i=1}^n(x_i-\overline{X})^2}\cdot\sqrt{\sum\limits_{i=1}^n(y_i-\overline{Y})^2}}
\end{eqnarray*}
where $\overline{X}$ and $\overline{Y}$ are the mean of each of the
random variables,  and $x_i$, $y_i$ are the $i$-th outcome of the
random variables $X$ and $Y$ respectively.

\end{document}